\documentclass[12pt,a4paper]{article}
\usepackage[utf8]{inputenc}
\usepackage[T1]{fontenc}
\usepackage{authblk}

\usepackage{nicematrix}
\usepackage{multirow}
\usepackage[table]{xcolor}

\usepackage{amsmath, amscd, amsthm, amsfonts, amssymb, graphicx, subcaption, color, soul, enumerate, xcolor,  mathrsfs, latexsym, bigstrut, framed, caption, listings}
\usepackage{pstricks,pst-node,pst-tree}
\usepackage[bookmarksnumbered, colorlinks, plainpages, backref]{hyperref}
\hypersetup{colorlinks=true,linkcolor=blue, anchorcolor=green, citecolor=midblue, urlcolor=darkblue, filecolor=magenta, pdftoolbar=true}

\usepackage[title]{appendix}
\usepackage{algorithm}
\usepackage{algpseudocode}

\usepackage{etoolbox}

\usepackage{setspace}

\usepackage{tikz}

\usetikzlibrary{shapes.geometric, arrows, positioning, fit, calc, backgrounds, arrows.meta, plotmarks}

\usepackage{pgfplots}
\pgfplotsset{compat=newest}

\usepackage[normalem]{ulem}

\newcommand{\Prob}{\ensuremath{\operatorname{Pr}}}

\definecolor{cupgreen}{rgb}{0,0.498,0.208}
\definecolor{cupblue}{rgb}{0,0,.5}
\definecolor{cupred}{rgb}{1,0.04,0}
\definecolor{cuppink}{rgb}{0.925,0,0.545}
\definecolor{cupmagenta}{rgb}{0.624,0.161,0.424}
\definecolor{cupbrown}{rgb}{0.71,0.212,0.133}

\definecolor{vibtilg}{HTML}{00FF99}
\definecolor{C_ev}{HTML}{FF8C00}
\definecolor{C_ce}{HTML}{FF0000}
\definecolor{C_mm}{HTML}{802F00}
\definecolor{C_ce+ls}{HTML}{4169E1}
\definecolor{C_mmlmc}{HTML}{228B22}
\definecolor{C_mmrls}{HTML}{4B0082}
\definecolor{des}{HTML}{edeb77}
\definecolor{welch}{HTML}{8bf2f7}
\definecolor{wilcox}{HTML}{e6a2f2}

\definecolor{TITLE}{rgb}{0,0,0}
\definecolor{midblue}{rgb}{0.00,0.0,0.80}
\definecolor{darkblue}{rgb}{0.00,0.00,0.45}
\definecolor{SECTION}{rgb}{0.50,0.00,1.00}
\definecolor{THM}{rgb}{0.8,0,0.1}
\definecolor{SEC}{rgb}{0,0,1}
\definecolor{calpolypomonagreen}{rgb}{0.12, 0.3, 0.17}
\definecolor{byzantine}{rgb}{0.64, 0.14, 0.54}
\definecolor{cadmiumgreen}{rgb}{0.0, 0.42, 0.24}
\definecolor{caribbeangreen}{rgb}{0.0, 0.6, 0.46}

\makeatother
\normalsize
\newtheorem{theorem}{{\color{THM} Theorem}}[section]

\theoremstyle{definition}
\newtheorem{definition}[theorem]{{\color{THM}Definition\ }}

\numberwithin{equation}{section}
\newtheorem*{definition*}{\color{THM}Definition}
\newtheorem*{observation*}{\color{THM}Observation}

\date{}
    \title{Overcoming Tight Constraints in Soft Happy Colouring}
    
    \author[1]{Mohammad H. Shekarriz \thanks{i.shekarriz@research.deakin.edu.au}}
    \affil[1]{School of Information Technology, Deakin University, Geelong, VIC, Australia}
    \author[1]{Asef 
Nazari\thanks{asef.nazari@deakin.edu.au}}
    \author[1]{Dhananjay Thiruvady \thanks{dhananjay.thiruvady@deakin.edu.au}}

\begin{document}

\maketitle
    
\begin{abstract}
The Soft Happy Colouring (SHC) problem, a mathematical framework for identifying homophilic network structures, seeks to maximise the number of $\rho$-happy vertices, i.e., vertices with at least a proportion $\rho$ of neighbours that share the same colour. Because this NP-hard problem makes finding exact solutions intractable for large networks, probabilistic metaheuristics such as the Cross-Entropy (CE) method are suitable candidates. However, pure CE frequently suffers from stagnation of the probability distributions and non-convergence in high-dimensional spaces. To address this, we introduce {\sf CE+LS}, synergising CE's adaptive learning with a fast, structure-aware local search ({\sf LS}). By restricting the search exclusively to local optima, 
{\sf CE+LS} learns from high-quality structural characteristics rather than raw random samples. We mathematically 
and empirically demonstrate that this search space reduction resolves CE's stagnation, yielding a convergent algorithm. Evaluating {\sf CE+LS} across 28,000 Stochastic Block Model graphs, validated by non-parametric statistical testing, demonstrates that it consistently outperforms existing heuristic and memetic algorithms. Furthermore, benchmarking against the commercial exact solver, CPLEX, on real-world networks confirms that {\sf CE+LS} identifies near-optimal configurations in a fraction of the required computational time for CPLEX. Crucially, {\sf CE+LS} remains highly efficient even in the tight constraint regime, where comparative algorithms usually fail.

\end{abstract}

\section{Introduction}\label{Sec:Intro}

\emph{Homophily} dictates the structural organisation of complex networks, manifesting as functional modules in biological systems, social circles in human interactions, and distinct communities across digital ecosystems~\cite{Homophily}. An effective approach to unravelling these structures is through the lens of graph colouring, specifically the Soft Happy Colouring (SHC) problem. 

The SHC problem is recognised as a graph colouring challenge. Instead of strictly requiring that all adjacent vertices have different 
colours, the goal is to maximise the number of $\rho$-happy vertices for $0\le\rho\le 1$, where a vertex is $\rho$-happy if the proportion of neighbouring vertices that share its colour is at least $\rho$~\cite{ZHANG2015117}. The motivation for introducing SHC was homophily in social networks that can be also expressed by the \emph{community structure} of graphs. A \emph{community}~\cite{10.1007/978-3-540-48413-4_23} in a graph is a subset of the vertex set whose size is ``large enough'', and its vertices are more densely adjacent to themselves compared with the remaining vertices. It is well-known that real-world graphs almost always exhibit community structures, a phenomenon extensively documented across general complex systems~\cite{Cherifi2019}, social and biological networks~\cite{doi:10.1073/pnas.122653799}, and models of human interaction~\cite{PhysRevE.68.065103}.

The SHC problem is NP-hard; finding an optimal solution is computationally intractable~\cite{ZHANG2015117}. Consequently, the use of metaheuristics~\cite{Blum2003} becomes inevitable for obtaining sub-optimal solutions to the problem. However, designing effective metaheuristics is a non-trivial task, requiring careful formulation of global and local search strategies that align with both the topology of the feasible region and the mathematical structure of the problem.

Recent theoretical analyses of SHC for the graphs in the Stochastic Block Model (SBM) in~\cite{SHEKARRIZ2025106893}, and its continuation in~\cite{SHEKARRIZ_local_search}, have established critical bounds on the proportion of happiness, $\rho$, that govern the behaviour of $\rho$-happy colourings. As will be formally detailed in Section~\ref{sec:Prelimin}, there exist two specific theoretical thresholds, $\mu$ (see Equation~\ref{eq:mu}) and $\tilde{\xi}$ (see Equation~\ref{eq:xi_tilde}), which dictate the structural feasibility of the problem. Based on these bounds, the parameter $\rho$ can be naturally classified into three distinct analytical regimes of the SHC problem for the graphs in the SBM, in increasing difficulty order:
\begin{itemize}
\item {\bf Mild regime:} when $0\le \rho <\mu$,
\item {\bf Intermediate regime:} when $\mu \le \rho\le \tilde{\xi}$, and
\item {\bf Tight regime:} when $\tilde{\xi}<\rho\le 1$.
\end{itemize}

The approach we propose here fundamentally enhances the Cross-Entropy (CE) method~\cite{rubinstein1999cross} through systematic search space reduction, tailoring a linear-time and effective tailored local search technique. 
The CE method takes an unconventional approach to finding the best possible solution to a problem by reframing it as a probability distribution learning problem. Instead of trying to directly pinpoint the absolute maximum value, the method sets a high target score slightly below the optimal value. A randomly generated solution that reaches or exceeds this challenging target is treated as a ``rare event''. The algorithm's primary goal then becomes estimating the likelihood of this rare event occurring. To achieve this, it repeatedly generates batches of random solutions, evaluates them, and identifies the top performers. By studying the characteristics of these elite few, the method continuously updates and refines the probability rules it uses to generate the next batch. Originally developed to simulate and study highly unlikely occurrences \cite{rubinstein1999cross}, this iterative process of learning from success allows the CE method to efficiently find optimal or near-optimal solutions without needing to evaluate every possibility~\cite{de2005tutorial}.

The probabilistic foundation of the CE method offers a flexible and robust framework applicable to a wide range of optimisation problems. It has been successfully extended to combinatorial optimisation \cite{rubinstein2004cross}, continuous optimisation \cite{rubinstein1999cross}, and mixed-integer nonlinear programming (MINLP) problems~\cite{bayat2024drawdown}. The integration of local search within a cross-entropy framework has also been explored in other combinatorial domains; for instance, Wang et al. \cite{wang2025cross} recently demonstrated the efficacy of a hybrid CE and local search approach for routing problems. However, as we will detail in Section \ref{sec:ce+ls}, the structure of the SHC problem necessitates a different hybridisation strategy than those used in path-based or continuous frameworks.

The CE method is suited to the SHC problem due to the discrete nature of the solution space. In SHC, the objective is to determine a colour assignment for each vertex that maximises the total number of $\rho$-happy vertices. The finite set of available colours allows us to construct a direct probability distribution over the entire space of possible assignments. The CE method seamlessly leverages this domain-specific structure. In this formulation, the probability that any given vertex belongs to a particular colour class becomes a dynamic and learnable parameter. Through successive iterations of generating solutions and evaluating the elite performers (better performing solutions), the CE method progressively refines these probabilities, effectively guiding the search toward optimal or near-optimal network partitions.

However, employing the pure CE method for SHC is problematic due to the multiplicity of competing locally optimal solutions in high-dimensional discrete spaces. This causes the pure CE mechanism to suffer from probabilistic stagnation, preventing convergence to the global optimal solution. By applying a fast and effective local search algorithm, such as {\sf LS}~\cite{SHEKARRIZ_local_search}, to each generated sample, we strictly limit the search space. This allows the CE mechanism to learn exclusively from high-quality structural skeletons rather than raw random samples. 

In this paper, we design the algorithm {\sf Cross-Entropy Local Search (CE+LS)} besides {\sf CE}\footnote{We distinguish between the Cross-Entropy method, abbreviated as CE, and the algorithm tailored for SHC which is denoted by {\sf CE}.}. Crucially, we theoretically prove and empirically demonstrate that this search-space reduction resolves stagnation in pure CE, transforming it into a convergent method characterised by fast decay of the Kullback-Leibler (KL) divergence in {\sf CE+LS}. By testing them over a large set of 28,000 randomly generated partially coloured graphs in the SBM, validated via robust non-parametric statistical analyses, and benchmarking against an exact solver for two separate but equivalent Integer Linear Programming (ILP) formulations of SHC on standard real-world networks, we show the superiority of {\sf CE+LS} in maximising the number of $\rho$-happy vertices. This dominance of {\sf CE+LS} is particularly pronounced in the tight constraint regime in the large test. 

In Section~\ref{sec:Prelimin}, we present preliminaries to our technical discussions. This includes notations from graph theory, the SBM, theoretical results about SHC, and its known algorithms. To benchmark the computational intractability of the problem, Section~\ref{sec:MIP} formulates two ILP models for SHC. The CE method for SHC is introduced in Section~\ref{sec:method}, while Section~\ref{sec:expriment} gives details and results of our experimental tests. Section~\ref{sec:conc} concludes the paper with a flavour of possible future works.

\section{Preliminaries}\label{sec:Prelimin}

Throughout the paper, a graph $G$ means a simple finite graph whose vertex and edge sets can represent, so we write $G=(V(G), E(G))$~\cite{Chartrand-graphs_and_digraphs}. The numbers of vertices and edges are usually denoted by $n$ and $m$.

Graphs, especially large real-world ones, are usually assumed to be modelled by random graphs~\cite{Bollobas_2001}. Here, we use the Stochastic Block model (SBM)~\cite{HOLLAND1983109}, specifically focusing on its simplified formulation~\cite{JERRUM1998155}. Hence, by the SBM, we mean the probability space consisting of all graphs on $n$ vertices, with an assignment to $k$ vertex-disjoint communities. The probability of having an edge between two vertices of the same community is $p$, while two vertices of different communities are adjacent with the probability of $q$. The necessary assumption is that $q<p$ to ensure meaningful community structure. This SBM ensemble is denoted by $\mathcal{G}(n,k,p,q)$. 

We utilise the SBM to generate benchmark networks, as its planted communities facilitate rigorous theoretical and practical analysis. However, our primary objective is to maximise network homophily, expressible as the number of $\rho$-happy vertices, rather than to strictly recover these ground-truth partitions. Because the SBM is a stochastic generative model, its planted labels do not necessarily guarantee the maximum possible homophily; the random distribution of edges frequently produces alternative partitions that are structurally more cohesive. Consequently, a divergence from the SBM's ground truth is not an algorithmic failure, but rather a reflection of the algorithm successfully identifying these alternative, highly homophilic configurations inherent to the generated graphs.

SHC was introduced in 2015 by Zhang and Li~\cite{ZHANG2015117} as a subordinate problem to the problem of \emph{Happy Colouring}. They sought a vertex colouring with a maximum number of \emph{happy vertices}, those that have the same colour as their neighbours. In a connected graph with some vertices precoloured, finding a happy colouring is a difficult problem because, in large and/or dense graphs, finding even one single happy vertex can be a challenge\footnote{It must also be noted that conventional happy colouring is a special case of soft happy colouring (when $\rho=1$), and because $\tilde{\xi}\leq 1$, the findings of~\cite{SHEKARRIZ2025106893} and~\cite{SHEKARRIZ_local_search} affirm the remark of~\cite{Lewis2019265} that there is almost no ($1$-)happy vertex in dense or large graphs.}. For the papers verifying this challenge, see~\cite{Lewis2019265}, which used an Integer Program and Construct, Merge, Solve \& Adapt, \cite{Zhang2018}, which considered a randomised LP-rounding technique and a non-uniform approach,~\cite{thiruvady2020} and~\cite{LEWIS2021105114}~propose tabu search approaches, and~\cite{THIRUVADY2022101188}~investigated evolutionary algorithms and hybrids of metaheuristics and matheuristics.

Formal problem definition of SHC is presented in the following definition. Note that a partial $k$-colouring for a graph $G$ is a function $\sigma: S\longrightarrow \{1,\ldots, k\}$ where $S\subsetneq V(G)$. A complete $k$-colouring is a function $\tilde{\sigma}:V(G)\longrightarrow\{1, \ldots, k\}$, whose domain is the entire vertex set of $G$. A $k$-colouring extension of the partial colouring $\sigma$ is a complete colouring $\tilde{\sigma}$ such that $\tilde{\sigma}(v) = \sigma (v)$ for all $v\in S$. 

\begin{definition}[\cite{ZHANG2015117}]
Suppose that $\sigma$ is a partially $k$-colouring of the graph $G$, $k\geq 2$, and $\tilde{\sigma}$ is a $k$-colouring extension $\sigma$. Then a vertex $v$ is \emph{$\rho$-happy (by $\tilde{\sigma}$)} if at least $\lceil \rho \cdot \deg(v) \rceil$ of the neighbours of $v$ have the same colour as $\tilde{\sigma}(v)$. A complete vertex colouring is called a soft happy colouring (with $k$ colours) for $G$ if it has the maximum number of $\rho$-happy vertices among such $k$-colouring extensions of $\sigma$.
\end{definition}

In a problem instance, which is a partially coloured graph, the uncoloured vertices are called \emph{free vertices}. Colours of free vertices can be different from one solution to SHC to another, but all valid solutions must keep the colours of non-free vertices unchanged.

By $H_\rho (\sigma) $ we mean the number of $\rho$-happy vertices of a (partial or complete) colouring $\sigma$ of the graph $G$, while $\sigma\in H_\rho$ means that $\sigma$ is a complete $\rho$-happy colouring, that is, a colouring that makes all the vertices $\rho$-happy. The \emph{ratio of $\rho$-happy vertices} of $\sigma$ is $\alpha(\sigma) =\frac{H_\rho (\sigma)} {n}$.

SHC for graphs in the SBM was explored in~\cite{SHEKARRIZ2025106893}. For a graph $G \in \mathcal{G}(n,k,p,q)$ with parameters $n=|V(G)|$, $2 \le k$, $0 < q < p < 1$, $0 < \rho \le 1$, and $0 < \varepsilon < 1$, they established that the planted community structure induces a $\rho$-happy colouring with probability at least $(1-\varepsilon)^n$, provided the following inequality holds:
\begin{equation}\label{th_eq}
    q(k-1)(e^\rho -1) + p(e^\rho - e) < \frac{k}{n}\mathrm{ln}(\varepsilon).
\end{equation}
Furthermore, they defined a threshold $\xi$ as
\begin{equation}\label{eq:xi}
    \xi = \max\left\{\min\left\{\mathrm{ln}\left(\frac{\frac{k}{n}\mathrm{ln}(\varepsilon)+p e +(k-1)q}{p+(k-1)q}\right), \; \frac{p}{p+(k-1)q}\right\}, \;
0\right\},
\end{equation}
such that for any $\rho \le \xi$, the underlying communities of $G$ constitute a $\rho$-happy colouring with high probability.

It was further asserted in~\cite{SHEKARRIZ2025106893} and~\cite{SHEKARRIZ_local_search} that for $G \in \mathcal{G}(n,k,p,q)$ in the asymptotic limit (as $n \to \infty$), this threshold converges to
\begin{equation}\label{eq:xi_tilde}
    \tilde{\xi} = \lim_{n\to\infty} \xi = \frac{p}{p+(k-1)q}.
\end{equation}
Consequently, the probability $\Prob(\sigma \in H_\rho)$ approaches 1 for $\sigma$ being the colouring induced by the communities of $G$ and $0 \le \rho < \tilde{\xi}$. Whereas for $\rho>\tilde{\xi}$, the expected value of $\alpha(\sigma ) $ falls to 0 as $n$ goes to infinity, and as a result, $\Prob(\sigma \in H_\rho)$ also approaches 0 for $\rho > \tilde{\xi}$. These theoretical findings were also substantiated through experimental validation on extensive sets of randomly generated graph instances~\cite{SHEKARRIZ_local_search}.

Moreover, for sufficiently small values of $\rho$, a significant proportion of vertices may be $\rho$-happy even within a colouring that bears no relation to the underlying community structure. In the context of SBM graphs, $G \in \mathcal{G}(n,k,p,q)$, a lower bound for $\rho$ has been identified~\cite{SHEKARRIZ_local_search}, below which a complete $\rho$-happy colouring is not expected to correlate highly with the graph's communities. This threshold is
\begin{equation}\label{eq:mu}
    \mu = \frac{q}{p+(k-1)q}.
\end{equation}
When $0 \le \rho < \mu$, it is possible for a vertex to satisfy the $\rho$-happy condition even if its colour predominantly aligns with vertices from other communities. Consequently, while achieving soft happiness is less demanding in this regime, the resultant colour classes are unlikely to yield a high-quality community detection~\cite{SHEKARRIZ_local_search}. 

For graphs generated by the SBM, a monotonic relationship exists between the parameter $\rho$ and the quality of the resulting community detection. Specifically, given $\rho_1 < \rho_2$, the alignment of a complete $\rho_2$-happy colouring with the ground-truth communities is demonstrably higher than that of a complete $\rho_1$-happy colouring~\cite{SHEKARRIZ_local_search}. Moreover, it is within the specific range $\mu \le \rho \le \tilde{\xi}$ that the colour classes of a complete $\rho$-happy colouring are considered to effectively represent a community structure for an SBM graph~\cite{SHEKARRIZ_local_search}. Consequently, the three regimes mentioned in the introduction, namely mild ($0\leq \rho <\mu$), intermediate ($\mu\leq \rho\leq \tilde{\xi}$), and tight ($\tilde{\xi}<\rho\leq 1$), make sense for graphs in the SBM because the nature of addressing the problem of SHC highly changes from one regime to another~\cite{SHEKARRIZ_local_search}. 

Known heuristic algorithms for SHC are as follows: Zhang and Li~\cite{ZHANG2015117} introduced two heuristic algorithms for SHC, namely {\sf Greedy} and {\sf Growth}\footnote{All the algorithms for SHC can have the suffix {\sf -SoftMHV}, following the notion of~\cite{ZHANG2015117}.}. In~\cite{SHEKARRIZ2025106893}, two more heuristics, namely {\sf LMC} and {\sf NGC}, are introduced, while {\sf LMC} is shown to be a fast and reliable algorithm for SHC that has a high correlation with the graph's community structure.
Three local search algorithms, namely {\sf LS}, {\sf RLS} and {\sf ELS}, are devised in \cite{SHEKARRIZ_local_search}. Among them, {\sf LS} was demonstrated to be a fast and effective local search algorithm for SHC, which can be used not only as a heuristic but also as an improvement algorithm. The algorithms {\sf LMC} and {\sf LS} are not only linear-time in terms of the number of edges ($\mathcal{O}(m)$), but also employ stochasticity, and thus their outputs are different every time they are run.  

Among the heuristic algorithms proposed for the SHC problem, three are foundational for the present work, as other heuristics are deemed unsuitable for integration into a metaheuristic due to either high time costs or deterministic outputs. For a heuristic to be used in a metaheuristic design, it is essential to be fast; otherwise, it will consume most of the available time and leave very little for solution maturation (learning). And, it must involve some stochasticity within its procedure so that, given an input, its output is generally different each time it is run.

The first required heuristic is {\sf Local Maximal Colouring (LMC)}, introduced in~\cite{SHEKARRIZ2025106893}. This algorithm is noted for its computational efficiency, possessing a linear time complexity of $\mathcal{O}(m)$. Furthermore, its output solutions have been shown to demonstrate a high correlation with the graph's intrinsic community structure. The design of {\sf LMC} operates independently of the proportion of happiness $\rho$ and incorporates stochasticity. Its central loop involves iteratively selecting a free vertex $v$ at random from the intersection of uncoloured vertices and the neighbours of already coloured vertices. This vertex $v$ is then assigned the plurality colour (i.e., the one most frequent) within its neighbourhood, $N(v)$.

The second algorithm, {\sf Local Search (LS)}, is another linear-time ($\mathcal{O}(m)$) procedure presented  in~\cite{SHEKARRIZ_local_search} that functions as both a heuristic and an improvement method. The algorithm is initialised by copying the input colouring $\sigma$ to a working solution $\tilde{\sigma}$. It then populates a set $U$ with all free vertices that are currently $\rho$-unhappy. The algorithm proceeds by iterating through $U$ in a randomised order, examining each vertex $v \in U$. If the vertex's colour $\tilde{\sigma}(v)$ does not agree with the plurality colour $q$ in its neighbourhood $N(v)$, its colour is updated by setting $\tilde{\sigma}(v) = q$. This process constitutes a single pass over the set of unhappy vertices.

The third method is {\sf Repeated Local Search (RLS)}, a local search algorithm also introduced in~\cite{SHEKARRIZ_local_search}. {\sf RLS} is structurally similar to {\sf LS}, but with a key iterative distinction: upon the completion of a full pass through the set $U$, the set is refilled with all vertices that are currently $\rho$-unhappy, and the process repeats. This iterative refinement naturally results in a higher time complexity than the single-pass {\sf LS}. Consequently, its application is generally limited to problem instances where the {\sf LS} algorithm by itself cannot achieve a significant improvement.

Six additional evolutionary algorithms for the SHC problem were presented in~\cite{SHEKARRIZ_Evolutionary}. The first three are \emph{Genetic Algorithms (GAs)}, namely {\sf GA(Rnd)}, {\sf GA(LMC)}, and {\sf GA(LS)}. These algorithms adhere to the standard genetic process, comprising \emph{selection}, \emph{crossover}, \emph{mutation}, and \emph{population modification}, and are distinguished solely by their method of generating the initial population: randomly, via {\sf LMC}, or via {\sf LS}, respectively. The remaining three are \emph{Memetic Algorithms (MAs)}: {\sf MA(Rnd)}, {\sf MA(LMC)}, and {\sf MA+RLS(LS)}. Memetic algorithms augment the genetic framework by incorporating a local improvement step before the population modification phase. The parenthetical notation again denotes the initialisation method. For {\sf MA(Rnd)} and {\sf MA(LMC)}, this improvement step is executed using {\sf LS}. However, for {\sf MA+RLS(LS)}, which is itself initialised by {\sf LS}, the more intensive {\sf RLS} is employed for the improvement step. It was noted that while these algorithms are pairwise statistically different, a degree of dependence was observed between {\sf MA(Rnd)} and {\sf GA(LS)}~\cite{SHEKARRIZ_Evolutionary}.

According to the empirical evaluation reported in~\cite{SHEKARRIZ_Evolutionary}, {\sf MA+RLS(LS)} yielded the superior average ratio of $\rho$-happy vertices, followed in performance by {\sf MA(Rnd)} and {\sf MA(LMC)}. The {\sf MA(LMC)} algorithm was noted for demonstrating the most accurate community detection, a characteristic inherited from the {\sf LMC} initialisation procedure. In contrast, {\sf GA(Rnd)} exhibited markedly inferior performance and was not considered competitive with the other five algorithms.

In Section~\ref{sec:method} we introduce two further algorithms for SHC, designated {\sf CE} and {\sf CE+LS}. Based on their metaheuristic design, it was hypothesised that {\sf CE} would be comparable to {\sf GA(Rnd)}, and {\sf CE+LS} would be comparable to {\sf MA(Rnd)}. While the solution quality of {\sf CE} was observed to be marginally superior to that of {\sf GA(Rnd)}, we will demonstrate in Section~\ref{sec:expriment} that {\sf CE+LS} not only surpasses {\sf MA(Rnd)} but also outperforms the previous state-of-the-art, {\sf MA+RLS(LS)}. This result positions {\sf CE+LS} as the most effective metaheuristic for SHC developed to date.

\section{ILP models for soft happy colouring}\label{sec:MIP}

Two mixed-integer programs are already  defined for maximising the number of (ordinary) happy vertices~\cite{Lewis2019265}. Here, we present formulations suitable for maximising the number of $\rho$-happy vertices.

For the first model, $M_1$, integer variables $x_i \in \{1,\ldots,k\}$ denote the colour assigned to a vertex $i\in\{1,\ldots,n\}$. Variables $y_i \in \{0,1\}$ are defined to be 1 if the vertex $i$ is not $\rho$-happy and 0 otherwise. To linearize the comparison of colours, we introduce auxiliary binary variables $w_{ij} \in \{0,1\}$ for each edge $(i,j) \in E$, which take the value 1 if $x_i = x_j$. As before, $n$ and $k$ represent the number of vertices and colours, respectively, and $V'$ is the set of precoloured vertices. The model's objective is to:
\begin{align}
\label{model:M1}
M_1: \text{maximise} &  \quad n - \sum_{i=1}^n y_i \\
\text{subject to:}\quad 
& x_i = c(i) && \forall i \in V', \label{model:M1x} \\
& x_i - x_j \leq k(1 - w_{ij}) && \forall (i,j) \in E, \label{model:M1w1} \\
& x_j - x_i \leq k(1 - w_{ij}) && \forall (i,j) \in E, \label{model:M1w2} \\
& \sum_{j\in N(i)} w_{ij} \geq \lceil \rho \deg(i) \rceil - \deg(i)\,y_i 
&& \forall i \in \{1,\ldots,n\} \label{model:M1y} \\
& x_i \in \{1,\ldots,k\}, y_i \in \{0,1\} && \forall i \in \{1,\ldots,n\}\\
& w_{ij} \in \{0,1\},  && \forall (i,j) \in E.\nonumber
\end{align}

In this model, Constraint~\ref{model:M1x} assigns the precolourings. Constraints~\ref{model:M1w1} and \ref{model:M1w2} guarantee that if the solver sets $w_{ij} = 1$ to claim two vertices share a colour, their colour values $x_i$ and $x_j$ must be strictly equal. The final Constraint~\ref{model:M1y} dictates $\rho$-happiness: because the objective function seeks to push $y_i$ to 0, it is forced to set $y_i = 1$ only when the sum of same-coloured neighbours falls below the required threshold $\lceil \rho \deg(i) \rceil$.

The second model, $M_2$, uses binary variables $x_{ic} \in \{0,1\}$ which are 1 if vertex $i$ is assigned colour $c \in \{1,\ldots,k\}$, and 0 otherwise. Variables $y_i \in \{0,1\}$ remain 1 if $i$ is not $\rho$-happy. To represent shared colours linearly, we introduce binary variables $z_{ijc} \in \{0,1\}$ which take the value 1 if both endpoints of edge $ij$ share colour $c$. Therefore, the model's objective is to:
\begin{align}
\label{model:M2}
M_2: \text{maximise} &  \quad n - \sum_{i=1}^n y_i \\
\text{subject to:}\quad
& x_{ic} = 1 && \forall i \in V' \text{ where } c = c(i), \label{model:M2x} \\
& \sum_{c=1}^{k} x_{ic} = 1 && \forall i \in \{1,\ldots,n\}, \label{model:M2x-c} \\
& z_{ijc} \leq x_{ic} && \forall (i,j) \in E,\ \forall c \in \{1,\ldots,k\}, \label{model:M2z1} \\
& z_{ijc} \leq x_{jc} && \forall (i,j) \in E,\ \forall c \in \{1,\ldots,k\}, \label{model:M2z2} \\
& \sum_{j\in N(i)} \sum_{c=1}^k z_{ijc} \geq \lceil \rho \deg(i) \rceil - \deg(i)\,y_i 
&& \forall i \in \{1,\ldots,n\}. \label{model:M2y} \\
& x_{ic} \in \{0,1\}, y_i \in \{0,1\}, z_{ijc} \in \{0,1\} && \forall i,j \in \{1,\ldots,n\}, \forall c \in \{1,\ldots,k\}. \nonumber
\end{align}
In this model, Constraint~\ref{model:M2x-c} guarantees assigning exactly one colour to each vertex. Constraints~\ref{model:M2z1} and \ref{model:M2z2} ensure that $z_{ijc}$ can only be 1 if both adjacent vertices $i$ and $j$ are assigned colour $c$. Finally, Constraint~\ref{model:M2y} mirrors the logic of $M_1$, effectively forcing $y_i = 1$ if the total count of same-coloured neighbours is insufficient to achieve $\rho$-happiness.

\section{The algorithms}\label{sec:method}

In this section, we introduce two algorithms for SHC. First in Section~\ref{sec:ce}, we introduce the main algorithm {\sf CE} and its functions, then in Section~\ref{sec:ce+ls}, we introduce {\sf CE+LS} and in Section~\ref{sec:convergence} we explain why {\sf CE+LS} converges to high-quality solutions for the SHC problems.

\subsection{The {\sf CE} algorithm for SHC}\label{sec:ce}

To solve an optimisation problem using the CE method, the problem must first be reformulated as a probability estimation task. The approach begins with a parametric probability distribution, which is iteratively refined to improve the quality of feasible solutions. This distribution serves as the basis for generating random sample data. At each iteration, the parameters of the distribution are updated to produce increasingly better samples~\cite{de2005tutorial}.

The CE method fundamentally reformulates deterministic optimisation problems as the estimation of rare event probabilities. For example, consider the problem of maximising a real-valued function $f(X)$ on $\mathcal{X}$; $$\gamma^*=\max\limits_{X \in \mathcal{X}} f(X).$$ In this context, for a random variable $X$, the event $\{f(X)\ge \gamma\}$ is considered a rare event when $\gamma$ is sufficiently close to $\gamma^*$. The CE method seeks to estimate the probability $\mathbb{P}(f(X)\ge \gamma)$, thereby guiding the search towards optimal or near-optimal solutions. Originally developed as a simulation technique for rare-event probability estimation \cite{rubinstein1999cross}, the CE method employs adaptive importance sampling to iteratively refine its search distribution. 

The CE method begins with a parametric probability distribution over the feasible region, often a discrete uniform distribution in the case of integer programming, and updates its parameters based on elite samples from each iteration. In essence, starting from an initial distribution, the CE method progresses through a sequence of distributions to reach a degenerate distribution with probability of 1.0 at the optimal solution and 0.0 everywhere else. The Kullback-Leibler divergence is used to measure the distance between two consecutive distribution functions. This adaptive mechanism enables a balance between exploration of the search space and exploitation of promising regions~\cite{de2005tutorial}. 

To solve SHC using the CE method, we begin with a discrete uniform probability distribution, where each free vertex is randomly assigned one of the $k$ colours. This initial random assignment constitutes a single feasible sample point. By repeating this process according to the initial distribution, we generate the first population of feasible solutions. For each solution in the population, we evaluate the number of happy vertices and rank the solutions accordingly. The top-performing solutions (eg, the top 10\%) form the elite set. This elite set is then used to update the parameters of the probability distribution, increasing the likelihood of assigning the most promising colours to each vertex.
This is done by measuring how many elite members the vertex $v$ has the colour $i$, for each free vertex $v$ and colour $i$, and then dividing them by the number of elite samples to calculate the updated probabilities. The next population, generated from this refined distribution, is expected to better explore the promising regions of the feasible space, thereby improving the overall solution quality~\cite{de2005tutorial}.

Thus, the \hyperref[alg:ce_main]{\sf CE} algorithm (Algorithm~\ref{alg:ce_main}) begins by initialising the best-found colouring, $\tilde{\sigma}$, with the precoloured vertices in the set $V'$. In Lines 2 to 4, for every free vertex $v$, the algorithm initialises a probability vector $P_v$ of size $k$, representing the probability of $v$ being assigned each colour. Each entry of this vector is therefore uniformly initialised to be $\frac{1}{k}$, and, in Line~5, the set of all such vectors is stored in $Probs$. At the beginning, the probability of a vertex $v_j \in V$ to get the colour $x_i \in \mathcal{C}$ is defined as $$\mathbb{P}(\sigma(v_j)=x_i)=\frac{1}{|\mathcal{C}|}.$$ When an elite set is detected, the probability of a particular vertex accepting a particular colour is modified based on the most frequent colour in the elite set. Consider the elite set as $E=\{ \sigma^{(1)}, \ldots, \sigma^{(N_E)}\}$, where $\sigma^{(l)}=(\sigma^{(l)}_{1}, \ldots, \sigma^{(l)}_{n})$. The colours of $v_j \in V$ in the elite set is represented as $\{\sigma^{(1)}_{j}, \ldots, \sigma^{(N_E)}_{j} \}$, where $\sigma^{(l)}_{j}\in \mathcal{C}$. With this notation, the probability of a vertex $v_i$ getting the colour $x_j$ in the next population is defined as follows.
\begin{equation}\label{eq:p}
    \mathbb{P}(\sigma(v_j)=x_i)= \frac{\text{number of times colour $i$ appeared for node $v_j$}}{\text{the size of the elite set}}=\frac{\sum\limits_{l=1}^{N_E}I_{\{\mathbf{c}^{(l)}(v_j)=x_i\}}}{N_E},
\end{equation} 
where $N_E$ is the size of the elite set and $I_{ \{\mathcal{P}\} }$ is an indicator function that is 1 when the proposition $\mathcal{P}$ is true and 0 otherwise.

\begin{algorithm}
    \caption{{\sf Cross-Entropy (CE)} --- Main Algorithm}\label{alg:ce_main}
    \begin{flushleft} $\;$\\ \hspace*{\algorithmicindent}
        \textbf{Input:} $G$, $\sigma:V'\longrightarrow \{1,\ldots,k\}$, $\rho$, $Population\_ Size$, $Elite\_ Size$, $\beta$\\ \Comment{$V'$ is the set of precoloured vertices and $\beta=$ smoothening factor}\\
        \hspace*{\algorithmicindent} \textbf{Output:} $\tilde{\sigma}:V(G)\longrightarrow \{1,\ldots,k\}$ 
    \end{flushleft}
    \begin{algorithmic}[1]
     
   \State $\tilde{\sigma}\gets \sigma$ \Comment{$\tilde{\sigma}(v)=\sigma(v),\; \forall v\in V'$}
        
        \For{$v\in V\setminus V'$}
            \State $P_v \gets \left[\frac{1}{k},\ldots, \frac{1}{k} \right]$ \Comment{$|P_v|=k,\;
\forall v\in V\setminus V'$}
        \EndFor
        \State $Probs\gets \{P_v \;
: \; v\in V\setminus V' \}$

        \While{$Terminate\_ Condition \neq$ True}
            \State $P \gets \Call{Form{\_}Population}{\sigma, V\setminus V', Probs, Population\_ Size}$
            \State $Scores\gets \{H_\rho (\tau)\; :\; \tau\in P\}$ \Comment{Every $\tau$ is an extension of $\sigma$}
            \State $\tilde{\sigma}\gets \tau$ for $\tau\in P$ such that $H_\rho (\tau)=\max Score$
            \State $Elite \gets \Call{Select{\_}Elite}{P, Scores, Elite\_ Size}$
            \State $Probs \gets \Call{Update{\_}Probs}{Probs, Elit, \beta, V\setminus V', k}$
        \EndWhile
        \State \textbf{Return} $\tilde{\sigma}$
    \end{algorithmic}
\end{algorithm}

The algorithm then enters its main loop in Line 6, which continues until a termination condition is met. Such a condition is usually a time limit or the number of $\rho$-happy vertices of $\tilde{\sigma}$ reaches the possible maximum. In each iteration of the main loop, a new population $P$ of $Population{\_}Size$ candidate colourings is generated by the $\Call{Form{\_}Population}{}$ operator in Line~7, which randomly assigns colours to free vertices based on the current colour probability stored in $Probs$. In Line~8, each colouring $c \in P$ is then evaluated using the objective function $H_\rho(c)$, and these values are stored in the vector $Scores$. Then, in Line~9, the best-performing colouring from this population updates $\tilde{\sigma}$. An elite set, $Elite$, containing the top $Elite\_ Size$ colourings, is selected in Line~10 via the operator $\Call{Select{\_}Elite}{}$. This elite set is then passed to the $\Call{Update{\_}Probs}{}$ operator in Line~11, which adjusts the probability vectors in $Probs$ to make it more likely that solutions similar to the elite set will be generated in the next iteration of the main loop. Once the loop terminates, the algorithm returns $\tilde{\sigma}$ as the best solution found.

\begin{algorithm}
    \caption{{\sf Form\textunderscore Population} ($\sigma, Free\_ Vertices, Probs, Population\_ Size$) --- Forms randomly generated colouring for free vertices based on the colour probabilities of vertices} \label{alg:form_population}
    \begin{algorithmic}[1]
        
            \State $P \gets \emptyset$
            \For{$t=1, 2, \ldots, Population\_ Size$}
                \State $\sigma_t \gets \sigma$ \Comment{Start with precoloured vertices fixed}
    
            \For{$v\in Free\_ Vertices$}
                    \State $\sigma_t (v) \gets \Call{RandomChoice}{Probs[v]}$ \Comment{Sample from $v$'s distribution}
                \EndFor
                \State $P \gets P \cup \{\sigma_t\}$
            \EndFor
         
   \State \Return $P$
        
    \end{algorithmic}
\end{algorithm}

The operator \hyperref[alg:form_population]{$\Call{Form{\_}Population}{}$} is responsible for the crucial generation phase of the \hyperref[alg:ce_main]{\sf CE} algorithm, producing a new set of candidate solutions, $P$, of size $Population\_ Size$. It accepts the initial precolouring $\sigma$, the set of $Free\_ Vertices$, and the current probability distribution $Probs$ as inputs.
The operator iterates $Population\_ Size$ times, generating a complete colouring solution $\sigma_t$ in each iteration.
For every new solution in Line~3, the colours of the pre-coloured vertices (defined by $\sigma$) are kept fixed.
In Lines~4 to 6, for each vertex $v$ in the set of $Free\_ Vertices$, the operator determines its colour by probabilistically sampling from the corresponding distribution vector $Probs[v]$.
This stochastic process ensures that the generated population reflects the current elite samples, concentrating sampling effort on regions of the search space that have historically produced high-scoring solutions.
Once $Population\_ Size$ solutions have been created, the resulting population $P$ is returned to the main algorithm for their fitness evaluation.

\begin{algorithm}
    \caption{{\sf Update\textunderscore Probs}($Probs^{\text{old}}, Elite, \beta, Free\_ Vertices, k$) --- Updates the colour probability of Vertices with the smoothening factor $\beta$} \label{alg:update_probs}
    \begin{algorithmic}[1]
        
           \State $Probs^{\text{new}} \gets \emptyset$
            \For{$v \in Free\_ Vertices$}
                \State $P_v^{\text{old}} \gets Probs^{\text{old}}[v]$ \Comment{Get the old probability vector for $v$}
                \State $P_v^{\text{raw}} \gets \text{new list of } k \text{ zeros}$ \Comment{Initialise raw probability vector}

                \For{$j \in \{1, \ldots, k\}$}\Comment{Calculate new probabilities based on the elite set}
                    \State $Count_j \gets |\{\tau \in Elite \; : \; \tau(v) = j\}|$
                    \State $P_v^{\text{raw}}[j] \gets Count_j / |Elite|$
                \EndFor
                
                \State $P_v^{\text{new}} \gets (\beta \cdot P_v^{\text{raw}}) + ((1 - \beta) \cdot P_v^{\text{old}})$
            
\EndFor
            \State $Probs^{\text{new}}\gets \{P_v^{\text{new}} \; : \;
v\in Free\_ Vertices \}$
            \State \Return $Probs^{\text{new}}$
        
    \end{algorithmic}
\end{algorithm}

The \hyperref[alg:update_probs]{$\Call{Update{\_}Probs}{}$} operator is critical for the learning phase of the \hyperref[alg:ce_main]{\sf CE} algorithm, refining the probability distribution $Probs$ based on the performance of the elite solutions. It takes the previous distributions ($Probs^{\text{old}}$), the set of high-performing colourings ($Elite$), and the $\beta$ (smoothening factor) as input. After initialisation of $Probs^\text{new}$ in line~1, the main loop of the operator begins, which iterates over the number of free vertices. For each $v \in Free\_ Vertices$, it first initialises the variables $P_v^\text{old}$ and $P_v^\text{raw}$ in Lines~3 and 4, then in lines~5 to 8 calculates a $k$-vector of $\text{raw}$ probabilities, $P_v^{\text{raw}}$, that stores the likelihood of a colour $j$ is the proportion of elite colourings that assigned colour $j$ to $v$. This $\text{raw}$ vector is then smoothed in Line~9, to form an updated probability vector $P_v^{\text{new}}$. This is done by the formula $\beta \cdot P_v^{\text{raw}} + (1 - \beta) \cdot P_v^{\text{old}}$. This weighted average ensures that the distribution shifts towards successful colourings (due to $P_v^{\text{raw}}$), while retaining some diversity and stability from $P_v^{\text{old}}$, preventing premature convergence. The collection of these new probability vectors forms the $Probs^{\text{new}}$, which is returned in Line~13 to the main algorithm.

\subsection{{\sf CE+LS} framework}\label{sec:ce+ls}

The local search enhanced extension of {\sf CE}, namely {\sf CE+LS}\label{alg:CE+LS}, is similar to {\sf CE} but uses the operator \hyperref[alg:form_population_ls]{$\Call{Form{\_}Population{\_}LS}{}$} instead of \hyperref[alg:form_population]{$\Call{Form{\_}Population}{}$}. As detailed previously in Section \ref{sec:Prelimin}, the {\sf LS} operator is a fast, greedy $\mathcal{O}(m)$ heuristic introduced in \cite{SHEKARRIZ_local_search} that iteratively assigns unhappy vertices to the plurality colour of their neighbourhood. 

The only difference between these two operators is that after Line~7 of the operator\linebreak \hyperref[alg:form_population_ls]{$\Call{Form{\_}Population{\_}LS}{}$}, Where {\sf LS} is run on every single colouring in the generated population. Using this operator instead of {\sf Form\textunderscore Population} enables the {\sf CE+LS} to leverage the fast, reliable local search algorithm {\sf LS} and converge more quickly to a high-quality solution.

\begin{algorithm}
\caption{{\sf Form\textunderscore Population\textunderscore LS}($\sigma, Free\_ Vertices, Probs, Population\_ Size$) --- Forms a population of colouring solutions based on the output of {\sf LS} over a randomly generated colouring for free vertices colour probabilities of vertices} \label{alg:form_population_ls}
    \begin{algorithmic}[1]
        
            \State $P \gets \emptyset$
            \For{$t=1, 2, \ldots, Population\_ Size$}
                \State $\sigma_t \gets \sigma$ \Comment{Start 
with precoloured vertices fixed}
                \For{$v\in Free\_ Vertices$}
                    \State $\sigma_t (v) \gets \Call{Random{\_}Choice}{Probs[v]}$ \Comment{Sample from $v$'s distribution}
                \EndFor
                \State $\tilde{\sigma_t}\gets \Call{LS}{\sigma_t,\rho, k, V'}$ \Comment{$V'$ is the set of precoloured vertices}
               
 \State $P \gets P \cup \{\tilde{\sigma_t}\}$
            \EndFor
            \State \Return $P$
        
    \end{algorithmic}
\end{algorithm}

The efficacy of the CE method lies in its importance sampling capability, which directs the search mechanism towards promising regions of the solution space by minimising the Kullback-Leibler divergence between successive probability distributions. Unlike Genetic Algorithms, which rely on pairwise crossover operators that may disrupt beneficial partial structures (the ``building block hypothesis''), the CE method updates parameters based on global aggregate statistics from the elite population. This enables a more robust reproduction mechanism that preserves global structural trends in the data.

For the Soft Happy Colouring problem with $k$ colours, the size of the discrete solution space is $k^{|V \setminus V'|}$, where $V$ and $V'$ respectively stand for the vertex set and the precoloured vertices. In other words, $V\setminus V'$ is the set of free vertices. In such high-dimensional spaces, the landscape of the objective function $H_\rho$ is often rugged with many local optima. The standard CE method effectively explores this space by maintaining a probability distribution vector $P_v$ for each vertex. Early in the process, high entropy in $P_v$ facilitates broad exploration. As iterations proceed, the smoothing parameter balances exploration of historical distributions and exploitation of the most recent elite samples, allowing the distributions to gradually concentrate on high-quality colouring assignments.

However, pure probabilistic methods can suffer from ``stagnation'', where probabilities converge to suboptimal deterministic values (0 or 1) too early. The integration of \textsf{LS} also addresses this by limiting the search to locally optimum solutions; by applying \textsf{LS} to each sample, the CE mechanism effectively learns from \emph{local optima} rather than raw random samples. This synergy is critical for community detection: the CE component identifies the global ``skeleton'' of the community structure (the coarse distribution), while the \textsf{LS} component efficiently enhances the quality of the generated solutions to refine the boundaries of these communities.

This theoretical hybridisation is engineered into the system architecture illustrated in Figure \ref{fig:flowchart_cels}. The diagram delineates the algorithmic flow into two distinct functional blocks: the \textit{Cross-Entropy Framework}, responsible for the macro-level via adaptive sampling, and the \textit{Hybrid Improvement} module, which executes micro-level local search. As shown, the critical handover occurs when raw candidate solutions are passed to the Local Search algorithm, transforming them into locally optimised solutions before fitness evaluation. This ensures that the ``Learning Phase'' (probability update) is driven exclusively by the statistics of high-quality, locally refined optima, thereby accelerating convergence towards a valid community structure. 

\begin{figure}[htbp]
\centering
\begin{tikzpicture}[
    node distance=1.5cm and 2cm,
    auto,
    startstop/.style = {rectangle, rounded corners, minimum width=3cm, minimum height=1cm, align=center, draw=black, fill=red!10, thick},
    process/.style = {rectangle, minimum width=3.5cm, minimum height=1.5cm, align=center, draw=black, fill=blue!10, thick},
    process_ls/.style = {rectangle, minimum width=3.5cm, minimum height=1.5cm, align=center, draw=black, fill=green!20, thick, dashed},
    decision/.style = {diamond, minimum width=3cm, minimum height=1cm, align=center, draw=black, fill=orange!10, thick},
    arrow/.style = {thick,->,>=stealth},
    line/.style = {thick}
]

    \node (start) [startstop] {Start};
    
    \node (init) [process, left=of start] {Initialise with Uniform \\Probability Distribution};
    
    \node (sample) [process, below=of init] {\textbf{Global Exploration}\\ Adaptive Sampling\\ (Generate Population)};
    
    \node (ls) [process_ls, right=of sample, xshift=1cm] {\textbf{Local Search}\\ $\sigma \to \tilde{\sigma}$};
    
    \node (eval) [process, below=of ls] {Evaluate Fitness\\ $H_\rho(\tilde{\sigma})$};
    
    \node (elite) [process, below=of eval] {Select Elite Samples\\ (Top Performer Subset)};
    
    \node (update) [process, left=of elite, xshift=-1cm] {\textbf{Learning Phase}\\ Update Probabilities\\ via Elite Samples};
    
    \node (decide) [decision, below=of update, yshift=-0.5cm] {Termination\\ Condition?};
    
    \node (best) [startstop, right=of decide, xshift=0cm] {Return Best Solution};

    \draw [arrow] (start) -- (init);
    \draw [arrow] (init) -- (sample);
    
    \draw [arrow] (sample) -- node[anchor=south] {Candidates} (ls);
    \draw [arrow] (ls) -- node[anchor=west] {Locally Optimised} (eval);
    \draw [arrow] (eval) -- (elite);
    \draw [arrow] (elite) -- (update);
    \draw [arrow] (update) -- (decide);
    
    \draw [arrow] (decide.west) -- node[anchor=south] {No} ++(-1,0) |- (sample.west);
    
    \draw [arrow] (decide.east) -- node[anchor=south, xshift=-0.6cm] {Yes} (best);

    \begin{pgfonlayer}{background}
        \node [fit=(init) (sample) (update) (decide), fill=blue!5, rounded corners, draw=blue!20, dashed, yshift=0.4cm, inner sep=0.5cm, label={[anchor=north, blue!60, font=\bfseries]north:Cross-Entropy Framework}] (ce_group) {};
        
        \node [fit=(ls) (eval) (elite), fill=green!5, rounded corners, draw=green!20, dashed, yshift=0.4cm, inner sep=0.5cm, label={[anchor=north, green!40!black, font=\bfseries]north:Hybrid Improvement}] (ls_group) {};
    \end{pgfonlayer}

\end{tikzpicture}
\caption[Flowchart of hybrid method {\sf CE+LS}]{Flowchart of the \textsf{CE+LS} Algorithm. The diagram illustrates the synergy between the Global Exploration (Adaptive Sampling) of the Cross-Entropy method and the Local Exploitation provided by the Local Search improver.}
\label{fig:flowchart_cels}
\end{figure}

It is critical to explicitly differentiate our proposed framework from existing hybrid methodologies to highlight its methodological novelty. Previous hybrid CE architectures, such as \cite{wang2025cross}, typically use local search as a final polishing step: the algorithm generates a population of random solutions, evaluates their raw fitness, selects the elite subset, and only then applies local search to improve those specific elites. In this standard approach, the core CE learning mechanism is still evaluating and navigating the vast, noisy space of raw random assignments.

In contrast, our framework mathematically embeds the local search before the evaluation and selection phase. Because our specific {\sf LS} operator is an ultra-fast, linear-time $\mathcal{O}(m)$ heuristic, we apply it to the \textit{entire} generated population immediately. Consequently, the {\sf CE+LS} never evaluates a raw random solution. Instead, the search space is fundamentally reduced: the probability matrix is tasked with learning patterns whose image under {\sf LS} demonstrate fitter solutions. Shifting the CE learning mechanism from navigating a noisy, near-infinite random space to strictly searching the space of local optima is a distinct methodological innovation that makes {\sf CE+LS} convergent, as we discuss it in more depth in the following section.

\subsection{Convergence of {\sf CE+LS}}
\label{sec:convergence}

As said, the SHC problem is NP-hard, and thus finding a polynomial-time algorithm for the global optimum is unlikely. However, we can establish asymptotic convergence of the {\sf CE+LS} framework; integrating {\sf CE} with {\sf LS} essentially modifies the way the topological landscape is investigated by mapping raw random samples to local optima, thereby effectively reducing the search space.

Let $\mathcal{X} = \{1, \dots, k\}^{|V \setminus V'|}$ denote the discrete state space of all possible colour assignments for the free vertices. Let $f: \mathcal{X} \to \mathcal{X}_{LS}$ be the deterministic, surjective mapping function representing the local search procedure, where $\mathcal{X}_{LS} \subset \mathcal{X}$ is the subset of SHC local optima. 
The objective function evaluated by {\sf CE+LS} is therefore the composite function $H_{\rho}(f(X))$, where $X$ is a random vector drawn from the parameterised probability distribution $P$.

\begin{theorem}
Given a graph $G$ and a smoothening factor $\beta \in (0, 1)$, the sequence of marginal probability matrices $\{P_t\}_{t=0}^{\infty}$ generated by the {\sf CE+LS} algorithm asymptotically converges to a deterministic target distribution $\bar{P}$ over the discrete search space $\mathcal{X}$, such that 
$$\lim_{t\to\infty} D_{KL}(\bar{P} \parallel P_t) = 0.$$
\end{theorem}

\begin{proof}
Let $\Theta$ represent the parameter space of all valid $|V\setminus V'| \times k$ row-stochastic probability matrices, with every row's sum is $1$. At any iteration $t$, {\sf CE+LS} generates a population of samples based on $P_t \in \Theta$, evaluates them via the composite function $H_{\rho}(f(X))$, and updates the parameters using the top-performing elite subset $\mathcal{E}_t$.

Let $\gamma_t = \min_{\sigma \in \mathcal{E}_t} H_{\rho}(f(\sigma))$ denote the minimum fitness within the elite set at iteration $t$, and let $\gamma^*$ be the global maximum number of $\rho$-happy vertices in $G$. The selection mechanism of the CE method ensures that the sequence $\{\gamma_t\}$ is monotonically non-decreasing and bounded above by $\gamma^*$. 

The update rule is defined as $$P_{t+1} = \beta P_{t}^{raw} + (1-\beta) P_{t},$$ where $P_{t}^{raw}$ is the empirical probability matrix of $\mathcal{E}_t$. Because $P_0$ is initialised uniformly at $\frac{1}{k}$, and $\beta \in (0, 1)$, it follows by induction that $P_t(i, j) > 0$ for any $t<\infty$. Consequently, for any configuration $x \in \mathcal{X}$ we have $\Prob_t(X=x) > 0$, which implies the search process operates is an irreducible Markov chain on the finite state space $\mathcal{X}$. 

Due to irreducibility and the finiteness of $\mathcal{X}$, the global optimum is almost surely sampled as $t \to \infty$. By the Monotone Convergence Theorem for bounded sequences, we have $\gamma_t \to \gamma^*$ almost surely. Because $\mathcal{X}$ is discrete, there almost surely exists a $T < \infty$ such that for all $t \ge T$, $\gamma_t = \gamma^*$. Consequently, for all $t \ge T$, the elite set $\mathcal{E}_t$ includes the global optima.

By restricting the search exclusively to the highly reduced subspace $\mathcal{X}_{LS}$, the local search aggressively homogenises the structural variance of the raw samples. As a result, for $t \ge t_0$, the empirical matrix $P_{t}^{raw}$ derived from the locally optimised elite sets stabilises towards a deterministic target matrix $\bar{P} \in \Theta$. 

For $t \ge T$, the update equation becomes
$$ P_{t+1} = \beta \bar{P} + (1-\beta) P_{t},$$
which by subtracting $\bar{P}$ from both sides yields
$$ P_{t+1} - \bar{P} = (1-\beta)(P_{t} - \bar{P}).$$
By applying any induced matrix norm, we have 
$$ \|P_t - \bar{P}\| = (1-\beta)^{t-T} \|P_T - \bar{P}\|.$$
Since $0<1-\beta <1$, we must have
$$ \lim_{t \to \infty} \|P_t - \bar{P}\| = 0 \implies \lim_{t \to \infty} P_t = \bar{P}.$$
Finally, we evaluate the KL-divergence $D_{KL}(\bar{P} \parallel P_t) = \sum_{x \in \mathcal{X}} \bar{P}(x) \ln \frac{\bar{P}(x)}{P_t(x)}$. Because $P_t \to \bar{P}$ pointwise, and no entry of $P_t$ equals absolute zero (preventing singularities in the logarithm), the continuity of the divergence function guarantees
$$ \lim_{t \to \infty} D_{KL}(\bar{P} \parallel P_t) = 0, $$
which completes the proof.
\end{proof}

Figure~\ref{fig:3d} conceptually illustrates the search space reduction mechanism central to {\sf CE+LS}. The semi-transparent 3D surface represents a continuous projection of the discrete, rugged objective landscape of the SHC problem, where the vertical axis denotes solution fitness, measured by the number of $\rho$-happy vertices. The horizontal axes, $x_1$ and $x_2$, act as a lower-dimensional abstraction of the complete high-dimensional search space, conceptually representing the various colour assignments available to the free vertices (where spatial proximity indicates structural similarity between candidate colourings). The red dots signify the raw candidate solutions generated randomly by the CE method's adaptive probability distribution during the global exploration phase. Rather than evaluating these raw samples directly, the algorithm applies a fast local search mechanism, visualised by the trajectory lines ($f$). These paths demonstrate how the local search actively refines the raw samples, pushing them strictly uphill to the nearest local optima, denoted by the green stars. Consequently, the CE learning phase restricts its probability updates exclusively to these locally optimised solutions (the green stars), allowing the algorithm to bypass low-fitness valleys and effectively learn from high-quality augmented samples rather than raw random samples.

\begin{figure}
    \centering
    \begin{tikzpicture}[
    scale=0.8,
    font=\sffamily,
    declare function={
        f(\x,\y) = 2.0*exp(-((\x-5)^2+(\y-5)^2)/1.5) +
                   1.2*exp(-((\x-2)^2+(\y-2)^2)/1.5) +
                   1.5*exp(-((\x-8)^2+(\y-8)^2)/1.5) +
                   1.0*exp(-((\x-2)^2+(\y-8)^2)/1.5) +
                   1.3*exp(-((\x-8)^2+(\y-2)^2)/1.5) +
                   1.4*exp(-((\x-5)^2+(\y-2)^2)/1.5) +
                   0.9*exp(-((\x-2)^2+(\y-5)^2)/1.5) +
                   1.1*exp(-((\x-8)^2+(\y-5)^2)/1.5) +
                   1.6*exp(-((\x-5)^2+(\y-8)^2)/1.5);
    }
]
    \begin{axis}[
        width=16cm,
        height=12cm,
        view={-35}{45},
        grid=major,
        xlabel={Search Space Dimension $x_1$},
        ylabel={Search Space Dimension $x_2$},
        zlabel={Fitness $H_\rho(\sigma)$},
        zmin=0, zmax=2.5,
        xmin=0, xmax=10,
        ymin=0, ymax=10,
        xtick=\empty, ytick=\empty, ztick=\empty,
        colormap={bluesurface}{color(0cm)=(white); color(1cm)=(blue!70!black)},
        clip=false 
        ]

        \addplot3[
            surf,
            opacity=0.45,
            faceted color=blue!40!black,
            samples=45,
            domain=0:10,
            y domain=0:10,
            z buffer=sort
        ] {f(x,y)};

        \addplot3[only marks, mark=star, mark size=6, mark options={color=green!80!black, line width=1.5pt}] table[row sep=crcr, x=x, y=y, z expr={f(\thisrow{x},\thisrow{y}) + 0.05}] {
            x y \\
            5 5 \\
            2 2 \\
            8 8 \\
            2 8 \\
            8 2 \\
            5 2 \\
            2 5 \\
            8 5 \\
            5 8 \\
        };

        \addplot3[only marks, mark=*, mark size=2.5, color=red!80] table[row sep=crcr, x=x, y=y, z expr={f(\thisrow{x},\thisrow{y}) + 0.02}] {
            x y \\
            3.5 4.5 \\
            6.5 5.5 \\
            4.5 3.5 \\
            1.5 1.0 \\
            8.5 7.0 \\
            7.8 8.5 \\
            1.5 7.0 \\
            7.5 1.5 \\
            5.0 1.0 \\
            1.0 3.8 \\
            7.5 4.0 \\
            4.2 8.0 \\
        };

        \draw[-{Stealth[length=3mm]}, thick, red!80] (axis cs:3.5, 4.5, {f(3.5,4.5)+0.02}) -- (axis cs:5, 5, {f(5,5)+0.02});
        \draw[-{Stealth[length=3mm]}, thick, red!80] (axis cs:6.5, 5.5, {f(6.5,5.5)+0.02}) -- (axis cs:5, 5, {f(5,5)+0.02});
        \draw[-{Stealth[length=3mm]}, thick, red!80] (axis cs:4.5, 3.5, {f(4.5,3.5)+0.02}) -- (axis cs:5, 5, {f(5,5)+0.02});
        \draw[-{Stealth[length=3mm]}, thick, red!80] (axis cs:1.5, 1.0, {f(1.5,1.0)+0.02}) -- (axis cs:2, 2, {f(2,2)+0.02});
        \draw[-{Stealth[length=3mm]}, thick, red!80] (axis cs:8.5, 7.0, {f(8.5,7.0)+0.02}) -- (axis cs:8, 8, {f(8,8)+0.02});
        \draw[-{Stealth[length=3mm]}, thick, red!80] (axis cs:7.8, 8.5, {f(7.8,8.5)+0.02}) -- (axis cs:8, 8, {f(8,8)+0.02});
        \draw[-{Stealth[length=3mm]}, thick, red!80] (axis cs:1.5, 7.0, {f(1.5,7.0)+0.02}) -- (axis cs:2, 8, {f(2,8)+0.02});
        \draw[-{Stealth[length=3mm]}, thick, red!80] (axis cs:7.5, 1.5, {f(7.5,1.5)+0.02}) -- (axis cs:8, 2, {f(8,2)+0.02});
        \draw[-{Stealth[length=3mm]}, thick, red!80] (axis cs:5.0, 1.0, {f(5.0,1.0)+0.02}) -- (axis cs:5, 2, {f(5,2)+0.02});
        \draw[-{Stealth[length=3mm]}, thick, red!80] (axis cs:1.0, 3.8, {f(1.0,3.8)+0.02}) -- (axis cs:2, 5, {f(2,5)+0.02}) node[midway, left, xshift=-2pt, yshift=4pt] {$f$};
        \draw[-{Stealth[length=3mm]}, thick, red!80] (axis cs:7.5, 4.0, {f(7.5,4.0)+0.02}) -- (axis cs:8, 5, {f(8,5)+0.02});
        \draw[-{Stealth[length=3mm]}, thick, red!80] (axis cs:4.2, 8.0, {f(4.2,8.0)+0.02}) -- (axis cs:5, 8, {f(5,8)+0.02});

        \node[align=left, anchor=north east, draw=black!50, fill=white, rounded corners, inner sep=6pt, scale=0.9, fill opacity=0.9, text opacity=1] 
        at (rel axis cs: 1.9, 0.9) {
            Raw samples  (\textcolor{red!80}{\Large $\bullet$})\\
            Local Search operation (\textcolor{red!80}{\textbf{--\textgreater}})\\
            Local Optima (\textcolor{green!80!black}{\Large $\star$})
        };

    \end{axis}
\end{tikzpicture}
    \caption{The diagram illustrates a continuous 2D projection of the discrete $|V \setminus V'|$-dimensional search space $\mathcal{X}$, where spatial proximity represents structural similarity between candidate colourings. The algorithm {\sf LS} is a method of finding a local optimum near the sample solution.}
    \label{fig:3d}
\end{figure}

To rigorously evaluate the information-theoretic learning efficiency and convergence behaviour of the standard {\sf CE} versus the proposed {\sf CE+LS} framework, an experimental trial was conducted across six independent SBM graphs. The networks were specifically tested within the structural phase transition regime ($\rho \approx \frac{\tilde{\xi}-\mu}{2}$), and the middle point of the tight regime ($\rho \approx \frac{1- \tilde{\xi}}{2}$). 
Both algorithms {\sf CE} and {\sf CE+LS} were executed using identical parameter configurations (a population size of 20, an elite size of 0.15, and a smoothing factor of 0.1).

\begin{figure*}[htpb]
   \centering
    \includegraphics[width=\textwidth]{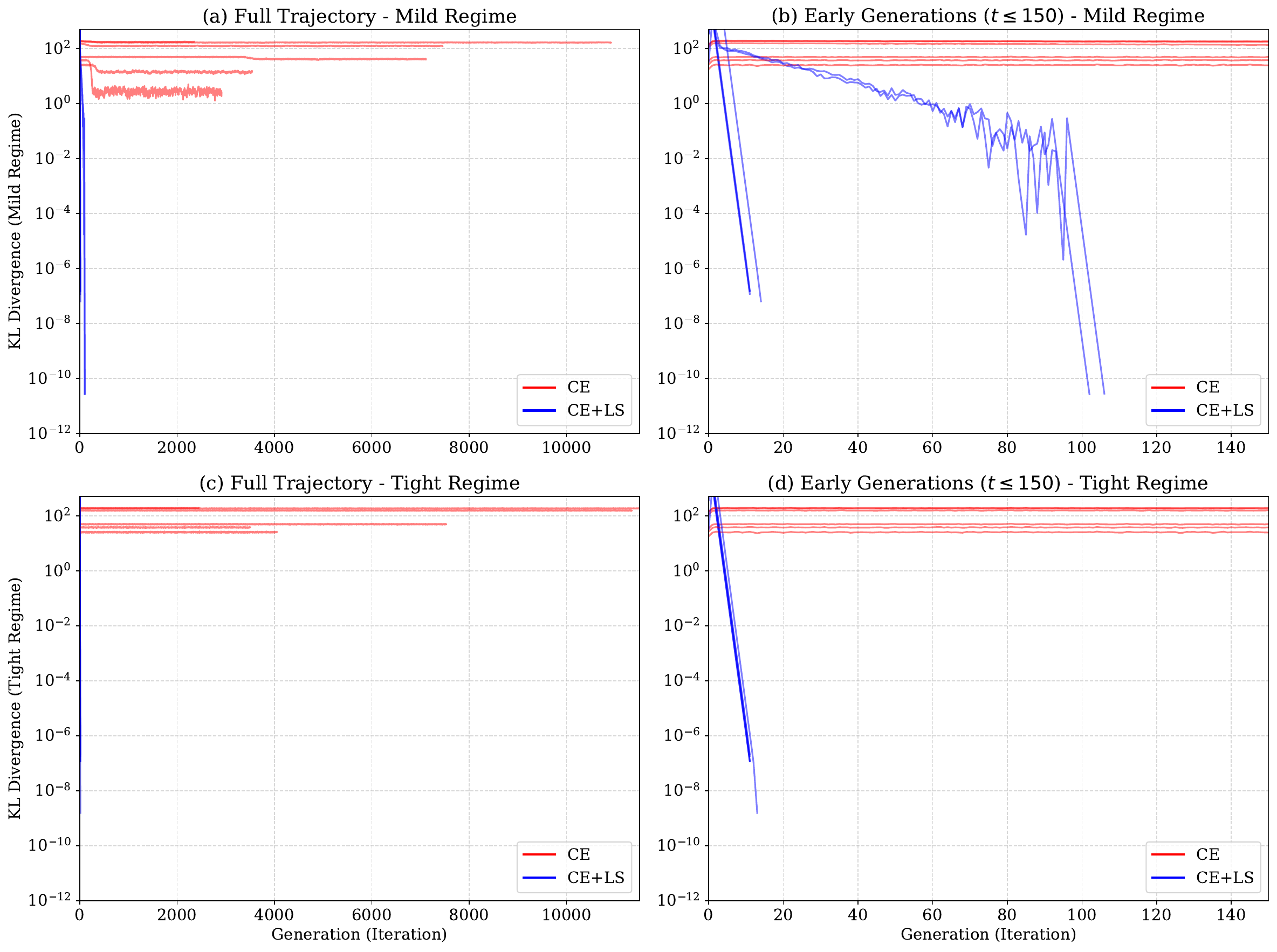}

    \caption{Empirical evolution of the KL-divergence evaluated within the mild regime (a and b) and within the tight regime (c and d).}
    \label{fig:kl_divergence}
\end{figure*}

The plotted results in Figure~\ref{fig:kl_divergence} trace the absolute Kullback-Leibler (KL) divergence between the algorithms' internal probability matrices and the ground-truth optimal configuration. By mapping the trajectories of the six individual networks concurrently, a stark mathematical contrast is revealed. The standard {\sf CE} algorithm exhibits immediate and permanent probability stagnation. Across all six independent instances, its KL-divergence metric forms a flat, non-convergent baseline (stagnating near $D_{KL} \approx 165$). Despite executing for over 10,000 generations, the pure {\sf CE} methodology demonstrates a complete inability to independently resolve the entropic uncertainty of the system's community boundaries.

Conversely, the {\sf CE+LS} algorithmic variant demonstrates a profound decay in KL-divergence. By integrating the local search component to actively repair deceptive, sub-optimal assignments within the candidate population, the framework acts as a structural oracle, rapidly dragging the cross-entropy updates toward higher objective values. The absolute KL-divergence for {\sf CE+LS} plummets monotonically across all tested graphs, consistently approaching zero (e.g., $D_{KL} < 10^{-10}$) within approximately 100 generations. This signifies a rapid, successful collapse of the probability matrix into a definitive, stable assignment, empirically proving that integrating local search is necessary to prevent premature probability convergence in highly constrained search spaces of SHC.

Furthermore, this accelerated convergence mechanism intrinsically resolves the numerical instability and stagnation vulnerabilities typical of pure CE methodologies. In a pure CE framework, unselected traits undergo geometric decay via the smoothing factor, requiring advanced sampling techniques (such as logarithmic scaling or Pareto sampling) to prevent machine-level floating-point underflow during extended iterations. However, because our full-population {\sf LS} operator definitively collapses the search space and achieves structural convergence within approximately 100 generations (as demonstrated in Figure 6), the absolute minimum probability value generated remains well within the safe operational limits of standard double-precision float architectures. Consequently, the hybridisation not only accelerates discovery but guarantees numerical stability without necessitating artificial probability inflation.

\section{Experimental results}\label{sec:expriment}

To compare {\sf CE} and {\sf CE+LS} with the existing algorithms of SHC, we perform practical tests over the set of 28,000 randomly generated precoloured graphs, which is introduced in~\cite{SHEKARRIZ2025106893} and used for the local search and evolutionary algorithms of~\cite{SHEKARRIZ_local_search, SHEKARRIZ_Evolutionary}.
The graph instances (stored in DIAMCS format) and algorithms source codes (in Python) are publicly available\footnote{at \href{https://github.com/mhshekarriz/HappyColouring_SBM}{https://github.com/mhshekarriz/HappyColouring\_SBM}}.
The graphs were defined for $200 \leq n < 3,000$ vertices.
For each $n$, 10 instances are generated, resulting in a total of 28,000 randomly generated graphs.
For each graph, parameters were randomly selected from the intervals as $k \in \{2,3,\ldots,20\}$, $p \in (0,1]$, $q \in (0, \frac{p}{2}]$, and $\rho \in (0,1]$. We set the time limit for our tests to 600 seconds to be able to compare the results with the evolutionary algorithms of~\cite{SHEKARRIZ_Evolutionary}.

For the remaining parameters of the {\sf CE} ($Population\_ Size$, $Elite\_ Size$, and $\beta$), a preliminary evaluation was performed using 1,000 randomly generated partially coloured graphs, constructed under identical conditions to the large-scale dataset. Empirical results indicated that the best average solution quality was obtained with $Population\_ Size = 20$, $Elite\_ Size = 0.15$, and $Smoothening\_ Factor = 0.1$. Accordingly, these settings were adopted for the comprehensive testing phase to ensure the consistency of results and enable direct comparison with existing benchmarks\footnote{The tests were performed via Deakin University's cloud computing server ``Gandalf'' over GPU nodes. The specifications are as follows: AMD EPYC 7402P CPU (24 cores / 48 threads) 256GB DDR4 RAM 4x Nvidia A100 GPUs (40Gb HBM2 Memory).}.

\subsection{The large test on randomly generated graphs in the SBM}\label{sec:test28000}

\begin{figure}
    \centering
    \includegraphics[scale=0.51]{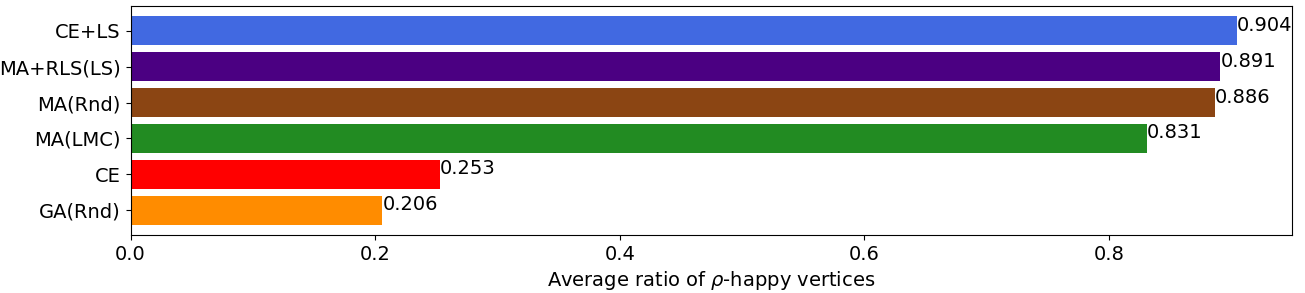}
    \caption[Average ratios of $\rho$-happy vertices of advanced algorithms]{Average ratios of $\rho$-happy vertices in the output of the tested algorithms when no condition is imposed on $\rho$.
}
    \label{fig:ce-happy-bar}
\end{figure}

First, we compare the algorithms for their average performance. Figure~\ref{fig:ce-happy-bar} presents a bar chart for the values of average $\alpha(\sigma)$ when $\sigma$ is a solution to the SHC problem for a graph in the set of tested 28,000 graphs. The {\sf CE+LS} algorithm demonstrates the best performance, with an average ratio of 0.904. 
Close behind are {\sf MA+RLS(LS)} at 0.891 and {\sf MA(Rnd)} at 0.886, suggesting that integration with the Local Search algorithm, {\sf LS}, is highly effective. In contrast, the {\sf MA(LMC)} algorithm shows a noticeably lower, but still much higher than the remaining two, with a ratio of 0.831. A significant drop in performance is observed for the algorithms without {\sf LS} components, with {\sf CE} achieving a low ratio of 0.253 and the simple {\sf GA(Rnd)} (Genetic Algorithm with Randomised population) performing the worst at 0.206.
This stark contrast highlights the importance of the local search method {\sf LS}, and the power of Cross-Entropy in the diversification of search.

To assess the statistical significance of the performance differences observed between the algorithms, we employed a dual-metric evaluation strategy combining both parametric and non-parametric methods. Table~\ref{table:stat-t-test} presents the consolidated matrix of pairwise $p$-values. The upper triangle of the matrix reports the results of Welch's $t$-test, which is robust against unequal variances across the massive sample distributions. To address the possibility that metaheuristic performance distributions may occasionally deviate from normality, the lower triangle reports the $p$-values calculated via the non-parametric Wilcoxon Signed-Rank test. The null hypothesis posits that there is no significant difference between the performance of any given pair of algorithms. The data emphatically reveals that for every pair of distinct algorithms, regardless of whether a parametric or non-parametric test is applied, the calculated $p$-value falls well below the standard significance threshold of $\alpha = 0.05$. Consequently, the null hypothesis is rejected in all cases. 

\begin{table}[htbp]
\small
\centering
\begin{NiceTabular}{|c|c|c|c|c|c|c|}[hvlines, code-before =
\rectanglecolor{vibtilg!50}{1-1}{2-1}
\rectanglecolor{C_ev!40}{3-1}{4-2}
\rectanglecolor{C_ev!40}{1-2}{2-2}
\rectanglecolor{welch!40}{3-3}{4-7}
\rectanglecolor{welch!40}{5-4}{6-7}
\rectanglecolor{welch!40}{7-5}{8-7}
\rectanglecolor{welch!40}{9-6}{10-7}
\rectanglecolor{welch!40}{11-7}{12-7}
\rectanglecolor{wilcox!40}{5-2}{14-2}
\rectanglecolor{wilcox!40}{7-3}{14-3}
\rectanglecolor{wilcox!40}{9-4}{14-4}
\rectanglecolor{wilcox!40}{11-5}{14-5}
\rectanglecolor{wilcox!40}{13-6}{14-6}
\rectanglecolor{C_ce!50}{5-1}{6-1}
\rectanglecolor{C_ce!50}{5-3}{6-3}
\rectanglecolor{C_ce!50}{1-3}{2-3}
\rectanglecolor{C_mm!35}{7-1}{8-1}
\rectanglecolor{C_mm!35}{7-4}{8-4}
\rectanglecolor{C_mm!35}{1-4}{2-4}
\rectanglecolor{C_mmlmc!40}{9-1}{10-1}
\rectanglecolor{C_mmlmc!40}{9-5}{10-5}
\rectanglecolor{C_mmlmc!40}{1-5}{2-5}
\rectanglecolor{C_mmrls!40}{11-1}{12-1}
\rectanglecolor{C_mmrls!40}{11-6}{12-6}
\rectanglecolor{C_mmrls!40}{1-6}{2-6}
\rectanglecolor{C_ce+ls!50}{13-1}{14-1}
\rectanglecolor{C_ce+ls!50}{13-7}{14-7}
\rectanglecolor{C_ce+ls!50}{1-7}{2-7}
]
\hline

              \Block{2-1}{ Algorithms} & \Block{2-1}{ \sf GA \\ (Rnd)} & \Block{2-1}{\sf CE} & \Block{2-1}{\sf MA \\ (Rnd)}
                & \Block{2-1}{\sf MA \\(LMC)} & \Block{2-1}{ \sf MA+RLS \\ (LS) }& \Block{2-1}{\sf CE+LS}\\ &&&&&&\\
\hline
\Block{2-1}{\sf GA(Rnd)}
              & \Block{2-1}{\textbf{---}}           & \Block{2-1}{0}        
    & \Block{2-1}{0}
               & \Block{2-1}{0}            & \Block{2-1}{0}            & \Block{2-1}{0}             \\&&&&&&\\
\hline
\Block{2-1}{\sf CE}
                 &\Block{2-1}{0}           & \Block{2-1}{\textbf{---}}          
  & \Block{2-1}{0} 
               & \Block{2-1}{0}            & \Block{2-1}{0}            & \Block{2-1}{0}             \\&&&&&&\\
\hline
\Block{2-1}{\sf MA(Rnd)}
               & \Block{2-1}{0}           & \Block{2-1}{0}            
& \Block{2-1}{\textbf{---}}
               & \Block{2-1}{$1.1\times 10^{-115}$}
                             & \Block{2-1}{0.022}            & \Block{2-1}{$8.7\times 10^{-21}$}            \\&&&&&&\\

\hline
\Block{2-1}{\sf MA(LMC)}
              & \Block{2-1}{0}          
 & \Block{2-1}{0}  
           & \Block{2-1}{$1.8\times 10^{-144}$}
               & \Block{2-1}{\textbf{---}}            & \Block{2-1}{$7\times 10^{-137}$}            & \Block{2-1}{$1.9\times 10^{-215}$}             \\&&&&&&\\
\hline
\Block{2-1}{\sf MA+RLS \\(LS)}
              & \Block{2-1}{0}          
 & \Block{2-1}{0}  
           & \Block{2-1}{$9.6\times 10^{-6}$}
               & \Block{2-1}{0}       & \Block{2-1}{\textbf{---}}            & \Block{2-1}{$2.1\times 10^{-12}$}             \\&&&&&&\\
\hline
\Block{2-1}{\sf CE+LS}
             & \Block{2-1}{0}           & \Block{2-1}{0}      
      & \Block{2-1}{0}
               & \Block{2-1}{0.000104}            & \Block{2-1}{0}            & \Block{2-1}{\textbf{---}}            \\&&&&&&\\
\end{NiceTabular}

\medskip
\caption{Comprehensive pairwise statistical significance matrix evaluating the algorithm performance differences. 
The upper triangle (cyan cells above the main diagonal) displays the $p$-values derived from Welch's $t$-test, while the lower triangle (pink cells below the main diagonal) displays the $p$-values derived from the non-parametric Wilcoxon Signed-Rank test. 
To simplify the visual presentation of the matrix, computed values falling below $1\times 10^{-250}$ are recorded as $0$.}
\label{table:stat-t-test}
\end{table}

It should be noted that the infinitesimally small $p$-values reported (e.g., $1.9 \times 10^{-215}$) are a direct mathematical consequence of the massive sample size ($28,000$ independent problem instances) evaluated in this study, which forces the standard error to approach zero and amplifies statistical certainty. Ultimately, both the Welch and Wilcoxon tests confirm that the performance profile of the proposed {\sf CE+LS} hybrid is statistically distinct from and superior to both the pure CE method and all competing memetic frameworks.

\begin{figure}
    \centering
    \includegraphics[scale=0.65]{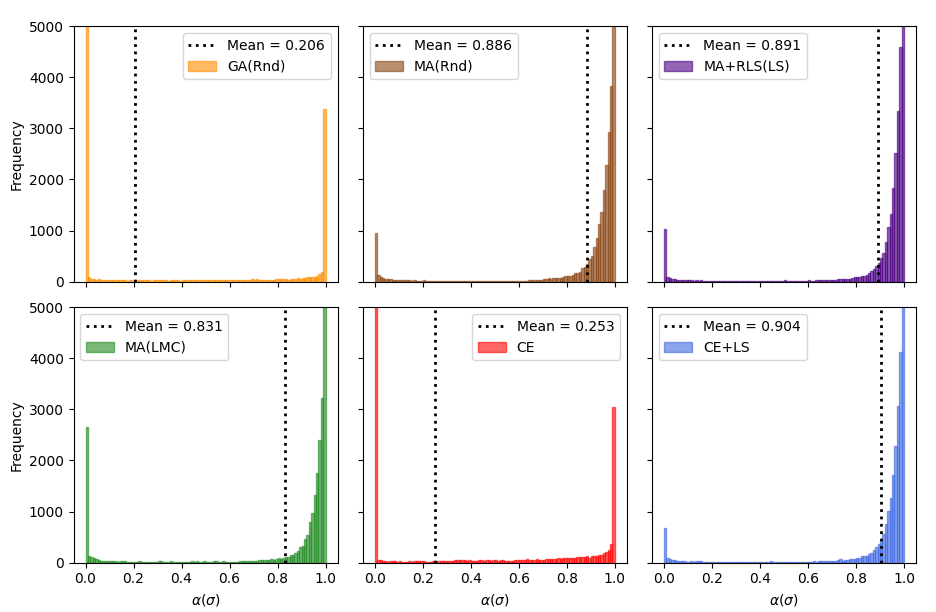}
    \caption[Histogram of ratios of $\rho$-happy vertices of advanced algorithms]{Histogram of ratios of $\rho$-happy vertices ($\alpha (\sigma)$) of colouring outputs ($\sigma$) of the tested algorithms.
For each of the six diagrams, the dotted vertical line represents the mean value that is also reported in Figure~\ref{fig:ce-happy-bar}.
The number of bins for demonstrating histograms is 100.
    }
    \label{fig:ce-hist}
\end{figure}

Figure~\ref{fig:ce-hist} consists of six histograms that illustrate the differential performance distributions of the algorithms with respect to the ratio of $\rho$-happy vertices, $\alpha(\sigma)$.
The highest-performing algorithms, {\sf CE+LS} (with a Mean of 0.904) and {\sf MA+RLS(LS)} (Mean 0.891), exhibit distributions that are highly concentrated and sharply peaked near $\alpha(\sigma)=1.0$, thereby demonstrating their superior solution quality and robust consistency.
A similar trend, though slightly less concentrated, is observed for {\sf MA(Rnd)} (Mean 0.886).
In contrast to {\sf CE+LS} and {\sf MA+RLS(LS)}, the {\sf MA(LMC)} algorithm (Mean 0.831) displays an almost similar distribution, with a higher peak near 0.0, indicating a lack of consistent convergence and occasional poor results that lower the average.
The algorithms with the lowest mean performance, {\sf CE} (Mean 0.253) and {\sf GA(Rnd)} (Mean 0.206), are characterised by a pronounced U-shaped or bimodal distribution, with high frequencies at both $\alpha(\sigma)\approx 0.0$ and $\alpha(\sigma)\approx 1.0$, though the latter is much less frequent.
This distributional profile is critical, as it confirms that their low mean performance is not due to moderate solution quality across all runs, but rather a consequence of a high frequency of near-complete failures being averaged with a minority of successful, near-optimal runs. This empirical behaviour is entirely consistent with the theoretical expectations established in Section~\ref{sec:convergence}. Without the structural guidance of local search, the pure CE method frequently succumbs to probabilistic stagnation, resulting precisely in the high volume of failed runs observed here.

\begin{figure}
    \centering
    \includegraphics[scale=0.45]{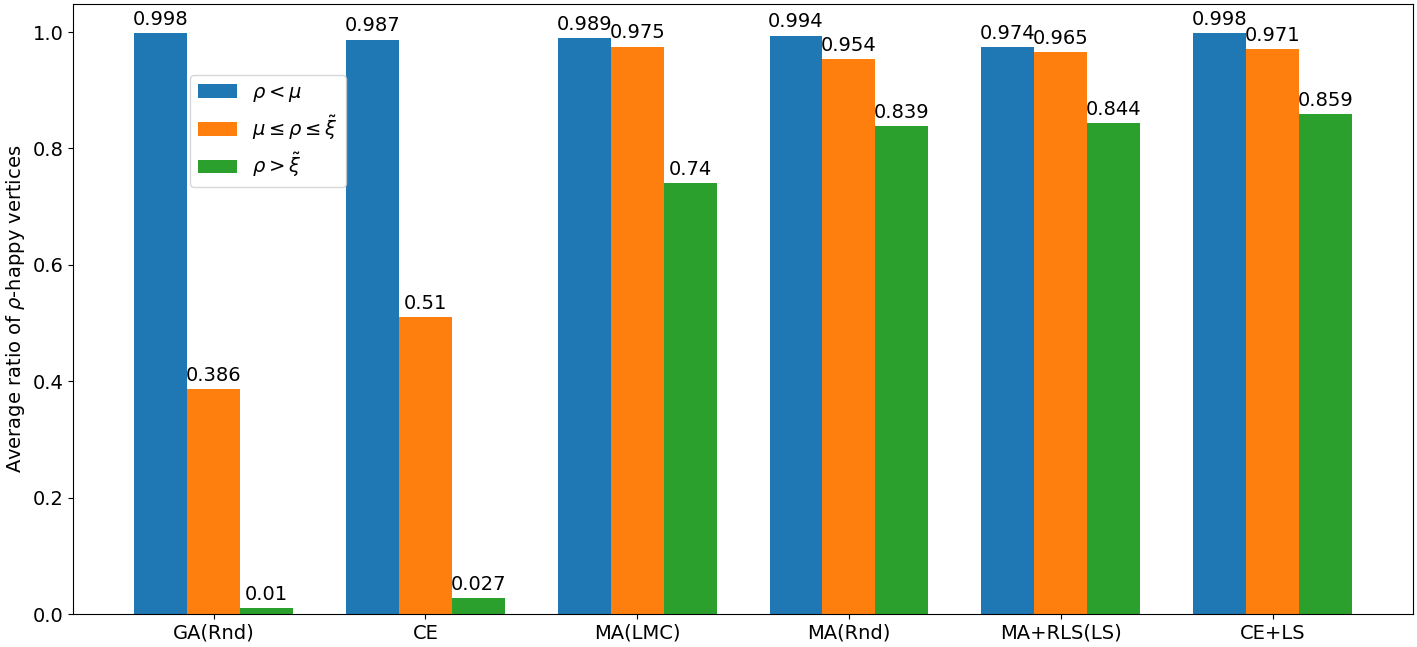}
    \caption[Average ratios of $\rho$-happiness in different regimes]{Average ratios of $\rho$-happy vertices in the output of the tested algorithms when $\rho<\mu$, $\mu\le\rho\le \tilde{\xi}$, or $\rho >\tilde{\xi}$.}
    \label{fig:ce-happy-cases}
\end{figure}

The clustered bar chart of Figure~\ref{fig:ce-happy-cases} dissects the average performance of the six algorithms by partitioning the ratio of $\rho$-happy vertices across the three distinct constraint 
regimes (mild: $\rho < \mu$, intermediate: $\mu \le \rho \le \tilde{\xi}$, and tight: $\rho > \tilde{\xi}$).
Under the mild regime, all algorithms perform nearly optimally, achieving ratios between $0.987$ and $0.998$.
However, the efficacy of the algorithms diverges sharply as constraints tighten.
In the intermediate 
regime, the local search ({\sf LS} or {\sf RLS}) based methods ({\sf MA(LMC), MA(Rnd), MA+RLS(LS), CE+LS}) maintain high performance (ratios $\ge 0.954$).
In contrast, the simpler {\sf GA(Rnd)} ($0.386$) and {\sf CE} ($0.51$) algorithms experience significant degradation.
This divergence is dramatically amplified in the tight constraint regime ($\rho > \tilde{\xi}$), where the most effective methods, particularly {\sf CE+LS} ($0.859$), {\sf MA+RLS(LS)} ($0.844$), and {\sf MA(Rnd)} ($0.839$), still yield substantial ratios, while {\sf GA(Rnd)} ($0.01$) and {\sf CE} ($0.027$) suffer near-total solution collapse.
These findings underscore that for highly constrained problem instances, the integration of the sophisticated {\sf LS} is a necessary condition for achieving high-quality solutions.

\begin{figure}
\captionsetup{size=small}
\begin{subfigure}{0.5\textwidth}
    \includegraphics[scale=0.47]{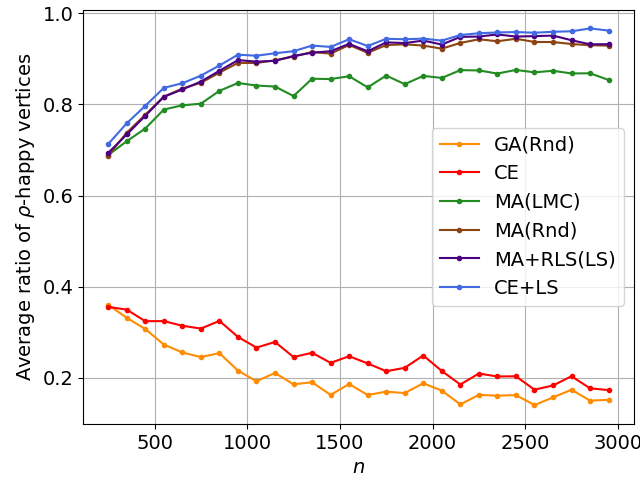}
    \caption{}\label{fig:ce-hap-n}
\end{subfigure} 
\begin{subfigure}{0.5\textwidth}
    \includegraphics[scale=0.47]{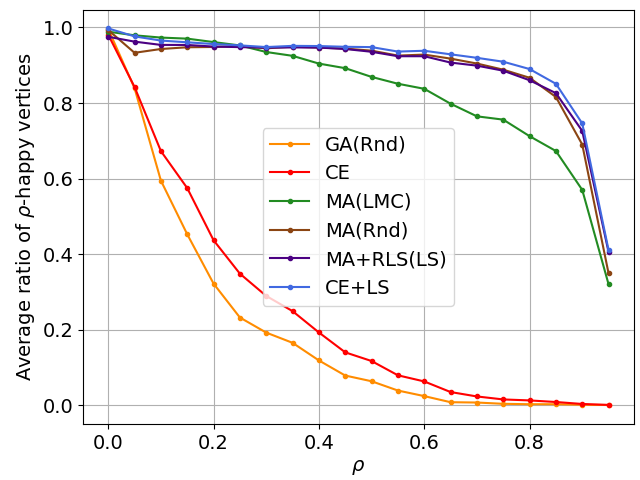}
    \caption{}\label{fig:ce-hap-r}
\end{subfigure} 

\begin{subfigure}{1\textwidth}
\centering
    \includegraphics[scale=0.47]{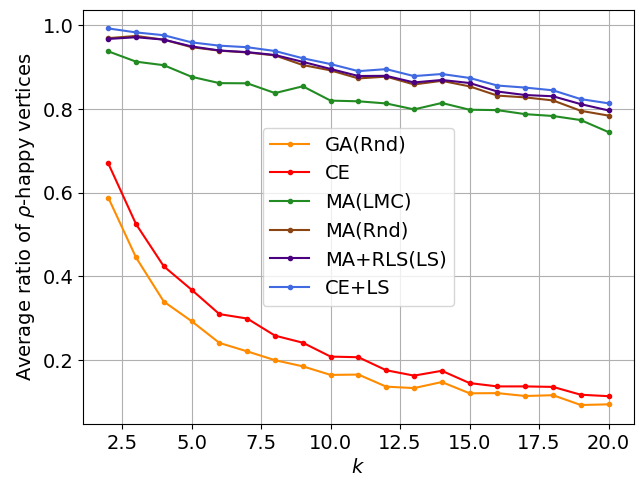}
    \caption{}\label{fig:ce-hap-k}
\end{subfigure} 

\caption[Average ratios of $\rho$-happy vertices wrt $n$, $\rho$, and $k$]{Comparison of the tested algorithms for their average ratios of $\rho$-happy vertices considering (a) the number of vertices $n$, (b) the proportion of happiness $\rho$, and (c) the number of colours $k$.}\label{fig:ce-hap}
\end{figure}

The plot in Figure~\ref{fig:ce-hap-n} illustrates the scalability of the six algorithms by tracking the average ratio of $\rho$-happy vertices as a function of the number of vertices, $n$. A significant differentiation in performance is immediately evident, separating the algorithms into two distinct groups.
The high-performing cluster, comprising {\sf  CE+LS}, {\sf  MA+RLS(LS)}, {\sf  MA(Rnd)}, and {\sf  MA(LMC)}, consistently achieves a high solution quality, with ratios rising from approximately $0.7$ at small $n$ to above $0.9$ for large $n$.
Notably, the three best-performing algorithms, {\sf  CE+LS}, {\sf  MA+RLS(LS)}, and {\sf  MA(Rnd)}, exhibit highly robust scalability, with their performance stabilising in the range of $0.92$ to $0.95$ as $n$ increases, confirming the efficacy of integrating local search.

In stark contrast, the lower-performing algorithms, {\sf  CE} and {\sf  GA(Rnd)}, demonstrate a marked negative correlation with problem size.
Their performance degrades monotonically as $n$ increases, dropping from initial ratios near $0.35$ to a plateau at approximately $0.20$ or below for $n \ge 1000$. This dependency on $n$ for the simpler methods indicates a severe lack of scalability, contrasting sharply with the size-independent near-optimality demonstrated by the hybrid metaheuristics.
The performance plot of Figure~\ref{fig:ce-hap-n} definitively establishes the consistent superiority of the {\sf  CE+LS}  across the entire range of graph sizes, $n$.

Although all advanced metaheuristics exhibit an improving trend in solution quality as $n$ increases, the {\sf  CE+LS} curve occupies the highest position at every measured point, confirming its status as the most effective solver for this optimisation problem.
Its average ratio of $\rho$-happy vertices stabilises above $0.95$ for larger problem instances, a level marginally, yet persistently, higher than its closest competitors, {\sf  MA+RLS(LS)} and {\sf  MA(Rnd)}.
This sustained dominance suggests that the specific synergy between the {\sf  CE}'s global search strategy and the powerful {\sf  LS} enhancement component yields the most potent and scalable optimisation mechanism, enabling it to consistently escape local optima and achieve the best-known solutions regardless of the initial problem scale.

The dependency of the average ratio of $\rho$-happy vertices on the proportion of happiness $\rho$ is illustrated in Figure~\ref{fig:ce-hap-r}.
All algorithms begin at near-perfect performance ($\approx 1.0$) for $\rho$ near $0$, but their resilience to increasing $\rho$ varies profoundly.
The algorithms are clearly segregated into two performance clusters. The superior cluster, consisting of the hybrid methods {\sf  CE+LS}, {\sf  MA+RLS(LS)}, {\sf  MA(Rnd)}, and {\sf  MA(LMC)}, maintains high solution quality ($\ge 0.9$) for $\rho$ values up to $\approx 0.6$, with only a significant drop occurring after $\rho=0.9$. Within this group, {\sf  CE+LS} and {\sf  MA+RLS(LS)} display the highest robustness, remaining above $0.8$ until $\rho \approx 0.8$. In stark contrast, the basic methods, {\sf  CE} and {\sf  GA(Rnd)}, exhibit an immediate and steep decay in performance: their ratio drops below $0.4$ for $\rho \ge 0.2$ and plummets to near $0$ for $\rho \ge 0.7$.
This strong inverse correlation for the simpler algorithms confirms that without the integrated power of local search, the ability to find $\rho$-happy vertices is rapidly diminished by even a moderate increase in $\rho$.

Figure~\ref{fig:ce-hap-k} examines the effect of the number of colours, $k$, on the average ratio of $\rho$-happy vertices. As the number of colours increases from $2$ to $20$, all algorithms experience a monotonic decline in performance, indicating that the problem becomes inherently more challenging with a larger colour palette.
The established dichotomy between the two algorithm clusters is consistently maintained across all values of $k$.
The advanced algorithms {\sf  CE+LS}, {\sf  MA+RLS(LS)}, {\sf  MA(Rnd)}, and {\sf  MA(LMC)} start near optimal performance ($\approx 0.98$) and robustly retain high solution qualities, with ratios stabilising above $0.8$ even for $k=20$.
Notably, the methods {\sf  CE+LS} and {\sf  MA+RLS(LS)} consistently occupy the top performance envelope.
Conversely, the basic algorithms, {\sf  CE} and {\sf  GA(Rnd)}, exhibit a sharp decay in performance, dropping from an initial ratio near $0.7$ for $k=2$ to below $0.2$ for $k=20$.
This pronounced inverse relationship demonstrates that the scalability and effectiveness of simple optimisation approaches are rapidly undermined by the increased complexity of the search space introduced by a larger number of colours.

\subsection{Validation on real-world graphs and exact solver benchmarking}

To verify the algorithmic efficacy of the {\sf CE+LS} framework on non-synthetic graphs, and to rigorously establish its necessity over exact mathematical solvers, an empirical evaluation was conducted on six standard real-world networks. Unlike synthetic Stochastic Block Models, real-world networks exhibit scale-free degree distributions characterised by highly connected hubs. 

The selected datasets represent a range of relatively small graphs, each possessing well-documented community boundaries used as baselines in network science literature. The networks used in this test are as follows:
\begin{enumerate}
    \item The Florentine Families dataset that charts the marital and business ties among Renaissance Florence's elite, highlighting a centralised power structure dominated by the Medici family \cite{padgett1993robust}.
    \item Zachary's Karate Club that maps the social interactions between members of a university karate club, famously capturing its factional split into two distinct groups \cite{zachary1977information}.
    \item The Dolphin Social Network that records the frequent associations between bottlenose dolphins in Doubtful Sound, New Zealand, demonstrating a natural biological division into two primary pods \cite{lusseau2003bottlenose}.
    \item The Les Mis\'{e}rables network that represents the co-appearances of characters within Victor Hugo's novel, forming highly clustered narrative sub-communities driven by major protagonists \cite{knuth1993stanford}.
    \item The Political Books (Polbooks) network that traces the co-purchasing behaviour of US political books during the 2004 presidential election, reflecting a stark partisan divide among readers \cite{krebs2004polbooks}.
    \item The American College Football network that maps the schedule of Division IA college football games during the 2000 regular season where the 12 ground-truth communities correspond perfectly to the official athletic conferences to which the teams belong \cite{girvan2002community}.
\end{enumerate}

The results, detailed in Table~\ref{tab:real_world_exact}, 
illustrate the limitations of commercial exact solvers (CPLEX) when confronted with scale-free networks at restrictive thresholds. For instance, determining the optimal bound for the Les Mis\'{e}rables network at $\rho=0.7$ required nearly 10 minutes of intense branch-and-bound exploration (546.4 seconds). Similarly, confirming the optimality of the Polbooks network at $\rho=0.9$ consumed 369.2 seconds. 

In contrast, the proposed {\sf CE+LS} metaheuristic, despite being constrained to only a 60-second time budget, demonstrated exceptional scalability and precision. In moderately constrained environments ($\rho=0.5$), the heuristic consistently identified the exact global optimum in less than one second. Crucially, even within the highly restricted landscapes ($\rho \in \{0.7, 0.9\}$) where exact solvers suffered from combinatorial explosion, {\sf CE+LS} achieved between 87\% and 100\% of the absolute mathematical maximum before terminating. 

These findings empirically confirm two primary strengths of the proposed framework. First, the cross-entropy probability distributions are capable of learning and adapting to complex structures without becoming trapped in local optima. Second, the hybrid metaheuristic approach is strictly necessary for large-scale applications of the Soft Happy Colouring problem, providing high-quality, near-optimal configurations in a fraction of the computational time required by exact methods.

\begin{table}[htpb]
    \centering
    
    \resizebox{\textwidth}{!}{
    \begin{tabular}{lccc|rr|rr|rr}
        \hline
        & & & & \multicolumn{2}{c|}{CPLEX (M1)} & \multicolumn{2}{c|}{CPLEX (M2)} & \multicolumn{2}{c}{{\sf CE+LS}} \\ \cline{5-10}
        Network & $n$ & $k$ & $\rho$ & $H_\rho (\sigma)$ & Time & $H_\rho (\sigma)$ & Time & $H_\rho (\sigma)$ & Time \\ \hline
        
\multirow{5}{*}{Florentine} & \multirow{5}{*}{15} & \multirow{5}{*}{3}
        & 0.10 & 15 & 0.05 & 15 & 0.03 & \textbf{15} & 0.03 \\
        & & & 0.30 & 15 & 0.03 & 15 & 0.03 & \textbf{15} & 0.05 \\
        & & & 0.50 & 15 & 0.34 & 15 & 0.41 & \textbf{15} & 0.05 \\
        & & & 0.70 & 9 & 21.27 & 9 & 0.68 & 8 & 60.00* \\
        & & & 0.90 & 8 & 19.98 & 8 & 0.04 & 7 & 60.00* \\ \hline
        
        \multirow{5}{*}{Karate Club} & \multirow{5}{*}{34} & \multirow{5}{*}{2}
        & 0.10 & 34 & 0.08 & 34 & 0.07 & \textbf{34} & 0.11 \\
        & & & 0.30 & 34 & 0.42 & 34 & 0.39 & \textbf{34} & 0.16 \\
        & & & 0.50 & 34 & 0.37 & 34 & 0.56 & \textbf{34} & 0.32 \\
        & & & 0.70 & 31 & 133.17 & 31 & 199.92 & \textbf{31} & 60.00* \\
        & & & 0.90 & 24 & 0.11 & 24 & 461.85 & 23 & 60.00* \\ \hline
        
        \multirow{5}{*}{Dolphins} & \multirow{5}{*}{62} & \multirow{5}{*}{2}
        & 0.10 & 62 & 0.41 & 62 & 0.76 & \textbf{62} & 0.23 \\
        & & & 0.30 & 62 & 0.57 & 62 & 1.16 & \textbf{62} & 0.20 \\
        & & & 0.50 & 62 & 0.78 & 62 & 0.65 & \textbf{62} & 0.78 \\
        & & & 0.70 & 62 & 0.68 & 62 & 0.96 & \textbf{62} & 0.42 \\
        & & & 0.90 & 62 & 0.46 & 62 & 0.87 & 55 & 60.00* \\ \hline
        
        \multirow{5}{*}{Les Mis\'{e}rables} & \multirow{5}{*}{77} & \multirow{5}{*}{5}
        & 0.10 & 77 & 16.21 & 77 & 0.99 & \textbf{77} & 0.33 \\
        & & & 0.30 & 77 & 318.00 & 77 & 1.02 & 75 & 60.00* \\
        & & & 0.50 & 76 & 368.96 & 76 & 258.49 & 74 & 60.00* \\
        & & & 0.70 & 68 & 502.95 & 68 & 546.43 & \textbf{68} & 60.00* \\
        & & & 0.90 & 51 & 78.48 & 51 & 24.57 & 46 & 60.00* \\ \hline
        
        \multirow{5}{*}{Polbooks} & \multirow{5}{*}{105} & \multirow{5}{*}{3}
        & 0.10 & 105 & 0.34 & 105 & 0.86 & \textbf{105} & 0.37 \\
        & & & 0.30 & 105 & 301.26 & 105 & 0.47 & \textbf{105} & 0.94 \\
        & & & 0.50 & 105 & 73.30 & 105 & 84.19 & 104 & 60.00* \\
        & & & 0.70 & 103 & 539.12 & 103 & 92.62 & 98 & 60.00* \\
        & & & 0.90 & 85 & 949.62 & 85 & 369.21 & 74 & 60.00* \\ \hline

        \multirow{5}{*}{Football} & \multirow{5}{*}{115} & \multirow{5}{*}{12}
        & 0.10 & 115 & 322.38 & 115 & 1.95 & 113 & 60.00* \\
        & & & 0.30 & 115 & 3720.75 & 115 & 88.45 & 108 & 60.00* \\
        & & & 0.50 & 111 & 98971.91 & 111 & 92734.70 & 103 & 60.00* \\
        & & & 0.70 & 104 & 99777.26 & 104 & 7636.32 & 77 & 60.00* \\
        & & & 0.90 & 73 & 1490.50 & 73 & 22038.14 & 27 & 60.00* \\ \hline
        
        \multicolumn{10}{l}{\footnotesize * Indicates the heuristic reached the strict 60-second computational time limit.}
    \end{tabular}
    }
    \caption{Comprehensive empirical evaluation on real-world networks across varying homophily thresholds ($\rho \in \{0.1, 0.3, 0.5, 0.7, 0.9\}$). The exact mathematical optimal bounds were derived using two commercial solver formulations: CPLEX M1 (integer) and CPLEX M2 (binary). The proposed {\sf CE+LS} metaheuristic was restricted to a strict 60-second single-threaded time budget. CPLEX runtimes are reported in cumulative CPU seconds across 32 threads.}
    \label{tab:real_world_exact}
\end{table}

To establish the scalability limits of the proposed framework, the American College Football network was included as a combinatorial stress test. Unlike the other networks, the Football dataset features 12 underlying communities ($k=12$), radically fragmenting the search space. For this network at $\rho=0.5$, the multi-threaded CPLEX solvers effectively exhausted a one-hour wall-clock limit, accumulating over 92,000 seconds of cumulative CPU time across 32 cores to confirm a maximum of 111 happy vertices. In stark contrast, the single-threaded {\sf CE+LS} algorithm achieved over 92\% of this mathematical upper bound (103 happy vertices) using only 60 seconds of CPU time, which represents a reduction in computational resource expenditure of over three orders of magnitude. 

However, as demonstrated in the $\rho \ge 0.7$ regimes for this specific 12-colour graph of the American College Football network, the combination of extreme homophily constraints and massive search-space fragmentation ultimately outpaces the aggressive 60-second heuristic budget, resulting in a degradation of solution quality. This boundary highlights that while {\sf CE+LS} is robust for standard partition sizes paired with extreme constraint thresholds may require an extended computational budget to allow the cross-entropy probability distributions sufficient generations to converge.

\subsection{The Necessity of Cross-Entropy Learning}

To definitively isolate the algorithmic contribution of the cross-entropy probability matrix, an ablation study was conducted across 1,000 highly heterogeneous SBM graphs\footnote{With the same parameter intervals as the main test.}. The proposed {\sf CE+LS} framework was evaluated against the memoryless {\sf Multi-Start Local Search (MSLS)} algorithm for SHC, which consists of a loop that starts {\sf LS} with random starting solution and records the colouring with the highest number of $\rho$-happy vertices. To ensure a rigorous comparison, both algorithms were subjected to a strict nominal 60-second computational budget.

\begin{figure}[htpb]
    \centering
\includegraphics[scale=0.45]{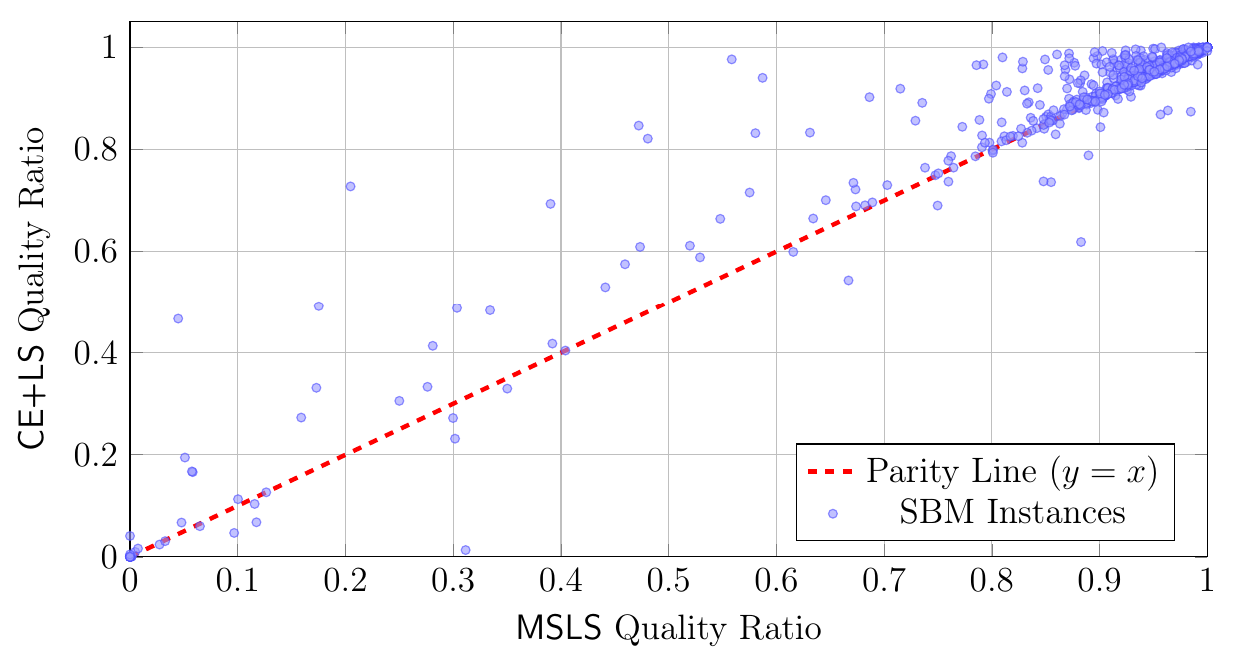}
    \caption{Performance scatter plot comparing the solution quality of {\sf CE+LS} against {\sf MSLS} across 1,000 distinct graph instances. Data points situated above the red parity line indicate instances where {\sf CE+LS} achieved a superior optimal bound.}
    \label{fig:ablation_scatter}
\end{figure}

As illustrated in the performance scatter plot (Figure~\ref{fig:ablation_scatter}), {\sf CE+LS} exhibited overwhelming stochastic dominance. Across the 1,000 instances, {\sf CE+LS} achieved a superior global bound in 63.9\% of instances, tied the baseline in 19.4\%, and was outperformed in only 16.7\%. Crucially, on instances where {\sf CE+LS} successfully bypassed the baseline, it recovered an average of 36.1 additional $\rho$-happy vertices per graph, visually represented by the heavy density of coordinates plotted above the parity line.

It is important to note a fundamental computational characteristic of population-based metaheuristics on large graphs ($|V| > 2000$). Because {\sf CE+LS} must evaluate an entire atomic generation prior to updating its probability matrix, executing a single initial generation on a large graphs can exceed more than 60 seconds. 
Conversely, the single-trajectory MSLS terminated precisely at the budget constraint. However, this temporal variance does not account for the performance disparity. When isolating the subset of smaller instances ($|V| \le 1000$) where both algorithms strictly adhered to the near-60-second budget, 
{\sf CE+LS} still maintained a dominant 87.5\% non-loss rate (129 wins, 61 ties, 27 losses). 

This extensive empirical analysis unequivocally confirms that the performance advantage of {\sf CE+LS} is derived fundamentally from its capacity for structural learning and intelligent search-space reduction, rather than extended computational time. The inclusion of the cross-entropy matrix is strictly necessary to prevent local search methodologies from being permanently deceived by complex community boundaries.

\subsection{Parameter calibration}\label{sec:calibration}

To ensure the integrity of the metaheuristic parameters and address the computational expense of applying local search to full populations, a formal calibration study was conducted prior to the main experiments. A representative sample of 1,000 graphs was evaluated with a 60 second time limit, generating random combinations of population sizes ($N$) and smoothing factors ($\beta \in (0, 0.20]$). The aggregated results are presented in Table~\ref{tab:calibration}.

\begin{table}[htpb]
    \centering
    \begin{subtable}{1\textwidth}
        \centering
        \caption{{\sf CE+LS}}
        \begin{tabular}{lcccc}
            \hline
            Population & \multicolumn{4}{c}{Smoothing Factor ($\beta$)} \\ \cline{2-5}
            Size ($N$) & 0.00--0.05 & 0.05--0.10 & 0.10--0.15 & 0.15--0.20 \\ \hline
            10--15 & 0.9612 & 0.9734 & 0.9552 & 0.9600 \\
            16--25 & 0.9562 & 0.9604 & 0.9576 & 0.9592 \\
            26--35 & 0.9574 & 0.9605 & 0.9516 & 0.9565 \\
            36--50 & 0.9321 & 0.9403 & 0.9401 & 0.9218 \\ \hline
        \end{tabular}

    \end{subtable}
    
    \vspace{0.5cm}
    
    \begin{subtable}{1\textwidth}
        \centering
        \caption{{\sf CE}}
        \begin{tabular}{lcccc}
            \hline
            Population & \multicolumn{4}{c}{Smoothing Factor ($\beta$)} \\ \cline{2-5}
            Size ($N$) & 0.00--0.05 & 0.05--0.10 & 0.10--0.15 & 0.15--0.20 \\ \hline
            10--15 & 0.3132 & 0.2860 & 0.3011 & 0.2451 \\
            16--25 & 0.2552 & 0.2552 & 0.2235 & 0.2372 \\
            26--35 & 0.2154 & 0.2406 & 0.2041 & 0.1878 \\
            36--50 & 0.2344 & 0.2006 & 0.1724 & 0.1935 \\ \hline
        \end{tabular}
    \end{subtable}

    \caption{Average ratio of happy vertices ($\alpha(\sigma)$) across 1,000 calibration graphs under a 60-second time limit, comparing the pure CE method against the hybrid {\sf CE+LS} framework.}
    \label{tab:calibration}
\end{table}

Table~\ref{tab:calibration} presents the parameter sensitivity analysis for the population size ($N$) and smoothing factor ($\beta$) under a fixed 60-second computational time budget. The empirical data demonstrates a strict performance trade-off governed by this time constraint. Highly restricted populations ($N < 15$) yield sub-optimal results due to inadequate statistical diversity in the elite subset, causing the probability matrix to converge prematurely before discovering deep local optima. 

Conversely, the data reveals a clear performance degradation when the population size exceeds $N=35$ (with average $\alpha(\sigma)$ scores dropping to $\approx 0.93$). This occurs because evaluating and applying the {\sf LS} operator to excessively large populations exhausts the strict operational time budget. Consequently, the algorithm is forced to terminate after completing only a fraction of the necessary global cross-entropy generations, preventing the probability matrix from achieving structural convergence.

Based on these empirical bounds, a population size of $N=20$ was selected as the optimal mathematical configuration for all primary experiments. This value balances sample diversity with generational throughput, yielding near-optimal homophily ($\alpha(\sigma) > 0.95$). Furthermore, a smoothing factor of $\beta=0.15$ was adopted; by preserving a historical component of the transition probabilities at each iteration.

\subsection{Time-to-Target analysis}

To evaluate the probabilistic variance and convergence behaviour of {\sf CE+LS}, a comparative Time-to-Target  analysis was conducted. A dense graph instance in the SBM was generated\footnote{With parameters $n=2876$, $k=14$, $p\approx 0.785$, $q\approx 0.039$, and $\rho=0.8$} and both {\sf CE+LS} and the {\sf MSLS} were executed for 100 independent runs. Crucially, to expose the exploration-exploitation dynamics, the algorithms were tasked with reaching an extremely constrained, globally optimal target configuration. 

\begin{figure}[htpb]
    \centering
\includegraphics[scale=0.45]{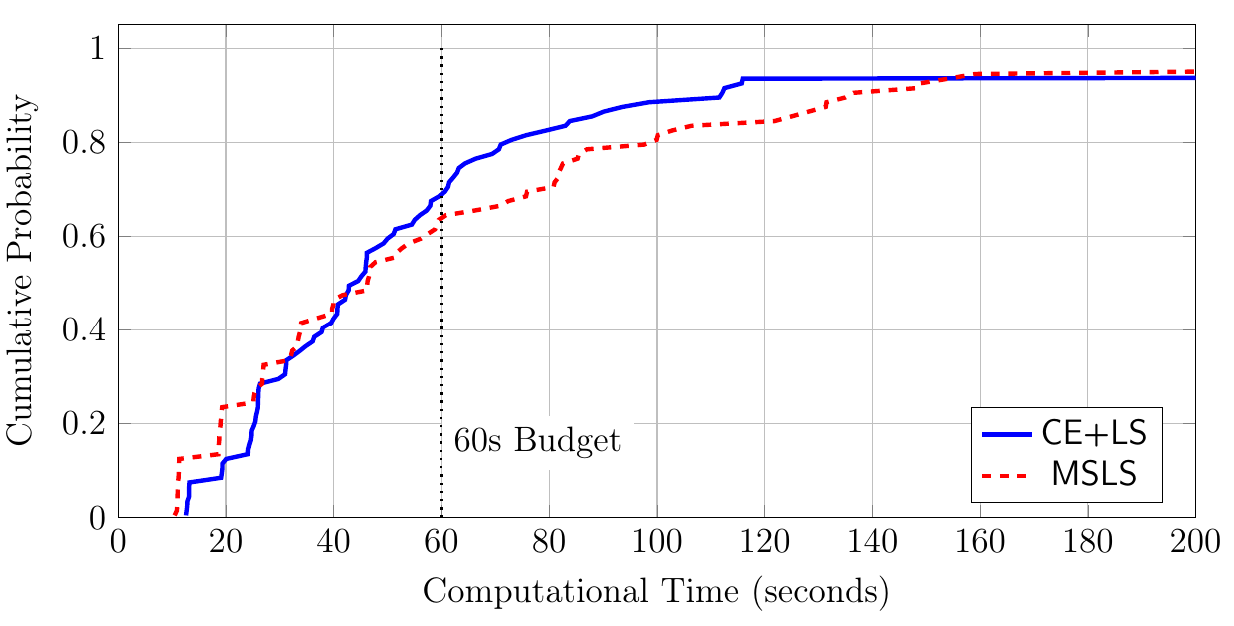}
    \caption{Comparative Time-to-Target plot evaluating the algorithms against a highly restrictive, near-global optimal target. Within practical computational time budgets (e.g., $t \le 120$ seconds), {\sf CE+LS} exhibits clear stochastic dominance, achieving significantly faster convergence. The heavy tail of  {\sf MSLS} requires extended, unconstrained computational time to match the success probabilities of the proposed framework.}
    \label{fig:ttt_comparative_plot}
\end{figure}

As depicted 
in Figure~\ref{fig:ttt_comparative_plot}, {\sf CE+LS} exhibits distinct stochastic dominance within practical operational time limits. By actively learning the topological structure of the community boundaries, the cross-entropy probability distributions significantly accelerate the local search procedure. At the 100-second mark, {\sf CE+LS} successfully reached the global optimum in 89\% of the independent executions, compared to only 81\% for the memoryless {\sf MSLS}. Furthermore, the 90th percentile convergence time ($Q_{0.90}$) for {\sf CE+LS} was achieved 23 seconds faster than {\sf MSLS} (111.6s vs. 135.1s).

However, the empirical data also highlights the classical theoretical trade-off inherent in model-based metaheuristics. In approximately 6\% of the independent runs, {\sf CE+LS} suffered from premature convergence, wherein the probability matrices converged toward deceptive local attractors, ultimately exhausting the strict 600-second time limit without escaping to the global summit. Conversely, because MSLS is entirely memoryless, it remains impervious to permanent deception, eventually locating the target in 100\% of its runs by executing thousands of brute-force random restarts over an extended timeframe (reaching up to 363 seconds). 

Ultimately, this analysis validates the operational design of the {\sf CE+LS} framework. While unconstrained, infinite-time scenarios can favour memoryless random restarts to guarantee global optimality, {\sf CE+LS} provides vastly superior, highly reliable convergence acceleration under the strict computational time budgets (e.g., $t \le 60$ seconds) required for practical, large-scale community detection.

\subsection{Accuracy of community detection}

As said in the introduction, the low ratio of the alignment of colour classes in an SHC problem with the underlying communities can cause no problem because SHC seeks homophily in graphs, as it may find an alternative significant partition equivalent to the graphs' original topology. However, to be consistent with prior works in the subject (i.e., ~\cite{SHEKARRIZ2025106893}, \cite{SHEKARRIZ_local_search}, and~\cite{SHEKARRIZ_Evolutionary}), we consider the data for this ratio in this section. 

\begin{figure}
    \centering
    \includegraphics[scale=0.51]{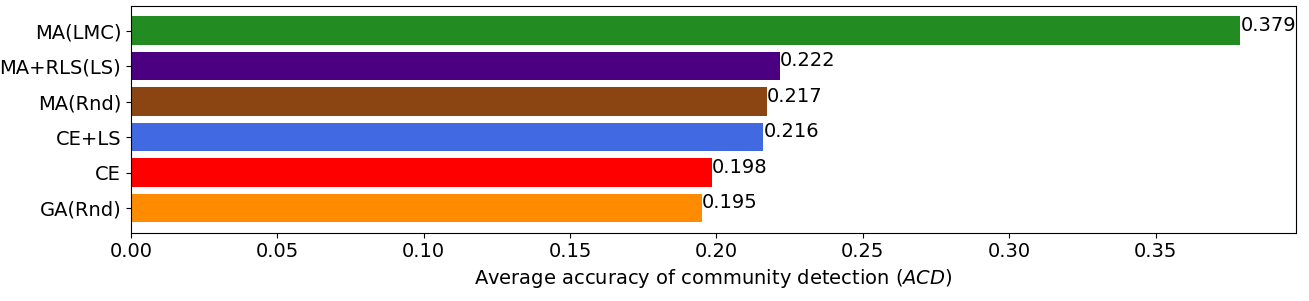}
    \caption[Average accuracy of community detection of advanced algorithms]{Average accuracy of community detection in the output of the tested algorithms when no condition is imposed on $\rho$.
}
    \label{fig:ce-comm-bar}
\end{figure}

The horizontal bar chart of Figure~\ref{fig:ce-comm-bar} presents a comparison of the Average Accuracy of Community Detection (ACD) achieved by the six algorithms.
In contrast to the previously analysed metrics, the {\sf MA(LMC)} algorithm demonstrates a substantial and singular lead, achieving an ACD of $0.379$.
This performance is approximately $70\%$ higher than the next closest competitor, indicating a significant and specialised advantage for the algorithms integrated with {\sf LMC} in maximising community detection accuracy.
The remaining five algorithms form a tight cluster of low performance, with their ACD values ranging narrowly from $0.195$ to $0.222$.
Within this lower group, the {\sf MA+RLS(LS)} algorithm records the highest value ($0.222$), followed closely by {\sf MA(Rnd)} ($0.217$) and {\sf CE+LS} ($0.216$).
The baseline algorithms, {\sf CE} ($0.198$) and {\sf GA(Rnd)} ($0.195$), exhibit the lowest accuracy.
This pronounced distribution highlights a critical distinction between the algorithms. While {\sf MA(LMC)} remains anchored to the planted SBM partition, {\sf CE+LS} successfully breaks away from the generative baseline to discover alternative partitions with higher internal homophily. This confirms that {\sf CE+LS} is not merely recovering a pre-defined structure, but is actively optimising the network topology to find the most cohesive organisation possible, which is often superior to the stochastic ground truth in the tight regime.

\begin{figure}
    \centering
    \includegraphics[scale=0.46]{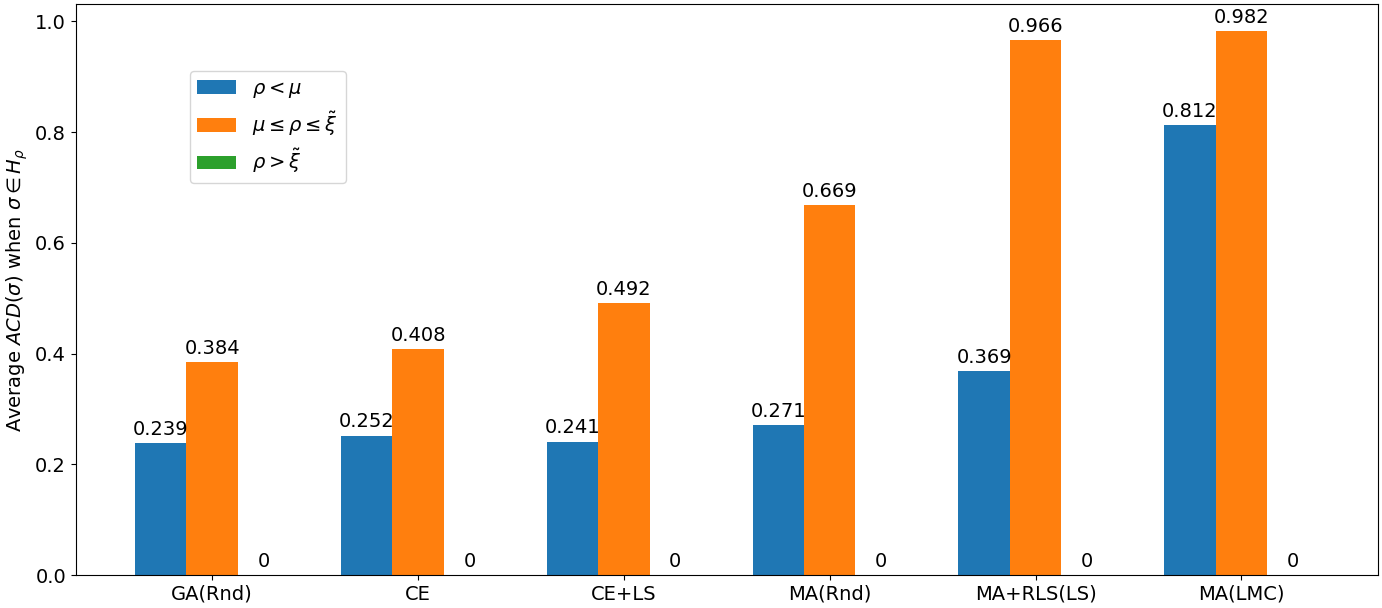}
    \caption[Average accuracy of community detection of complete solutions by advanced algorithms]{Average accuracy community detection ($ACD$) when the tested algorithms found complete $\rho$-happy colourings.
When $\rho >\tilde{\xi}$, no algorithm could find a complete $\rho$-happy colouring.
Consequently, no green bar is visible on the chart.}
    \label{fig:th-comm-hap-comp}
\end{figure}

This clustered bar chart of Figure~\ref{fig:th-comm-hap-comp} analyses the \emph{Average Accuracy of Community Detection ($\text{ACD}(\sigma)$)} conditional on the solution being a complete $\rho$-happy solution ($\sigma \in H_{\rho}$), segregated across the three constraint regimes (Mild: $\rho < \mu$, Intermediate: $\mu \le \rho \le \tilde{\xi}$, and Tight: $\rho > \tilde{\xi}$).
The data reveals a critical insight: in the tight regime, no algorithm can achieve solutions that are completely $\rho$-happy (i.e., when $\sigma \in H_{\rho}$).
Thus, the community detection accuracy vanishes (i.e. $\text{ACD}(\sigma)=0$) in the tight constraint regime of $\rho > \tilde{\xi}$, exactly as it was predicted by~\cite[Theorem 3.1]{SHEKARRIZ_local_search}.
For the two milder regimes ($\rho < \mu$ and $\mu \le \rho \le \tilde{\xi}$), the {\sf MA(LMC)} and {\sf MA+RLS(LS)} algorithms demonstrate outstanding, near-optimal performance, achieving the highest conditional ACD values, with {\sf MA(LMC)} peaking at $0.982$ (intermediate) and {\sf MA+RLS(LS)} at $0.966$ (intermediate).
This high conditional accuracy suggests that the structural properties enforced by a $\rho$-happy colouring are highly congruent with the actual community structure, provided that the colouring is obtained by an advanced, structure-aware algorithm, such as {\sf LMC}, integrated into the memetic strategies.
Conversely, the remaining four algorithms, including {\sf CE+LS}, exhibit significantly lower conditional ACD values, never surpassing $0.492$, indicating that while they can find $\rho$-happy solutions, those solutions can represent community structures different from the underlying community structure.

\section{Conclusion}\label{sec:conc}

We have advanced the algorithmic landscape for soft happy colouring (SHC) by engineering a mathematically convergent framework, {\sf Cross-Entropy Local Search (CE+LS)}. While the pure {\sf Cross-Entropy (CE)} method struggles with the multiplicity of optimal solutions, which results in stagnation of the probability distributions, we deliberately designed {\sf CE+LS} to overcome this by utilising a fast local search ({\sf LS}) over the entire generated population to strictly reduce the vast search space to structural local optima. This integration enforces strict agreement among global probability updates, ensuring that the algorithm is mathematically convergent. We have formally proved this convergence and empirically validated, across multiple independent network instances, that the absolute KL-divergence of {\sf CE+LS}'s iterations forms a convergent sequence that definitively and rapidly approaches zero.

To benchmark this engineered framework, we conducted extensive evaluations across 28,000 randomly generated partially coloured graphs. The results, validated by dual-metric statistical analyses, conclusively demonstrate that {\sf CE+LS} unequivocally outperforms every existing population-based evolutionary algorithm for SHC. Notably, {\sf CE+LS} establishes absolute dominance in the highly challenging tight constraint regime of $\tilde{\xi}<\rho\leq 1$. Furthermore, benchmarking against exact Integer Linear Programming formulations solved via a commercial solver (CPLEX) on real-world networks validated the necessity of this metaheuristic approach. The {\sf CE+LS} framework identified exact or near-optimal configurations within a strict 60-second budget, overcoming homophily constraints that caused commercial branch-and-bound solvers to become computationally intractable.

Crucially, our analysis of community detection explicitly distinguishes between optimising the SHC objective and recovering planted labels. While {\sf CE+LS} achieves unparalleled success in maximising homophily, we demonstrated that this strict optimisation frequently discovers alternative cohesive structures that diverge from, and are often denser than, the original generative SBM ground truth.

Future research can consist of many directions. First, applying the {\sf CE+LS} framework to massive-scale real-world networks, such as biological protein-protein interaction networks or dynamic social media graphs, would provide valuable insights into its practical robustness in detecting latent homophilic clusters under extreme noise. Second, the potential of integrating more sophisticated local search mechanisms remains unexplored; synergising the Cross-Entropy method with Tabu Search or Variable Neighbourhood Search could offer superior diversification strategies to escape deep local optima. Additionally, investigating adaptive parameter control mechanisms for the smoothing factor and elite sample size could further automate the solver, reducing the dependency on manual calibration. Finally, extending this hybrid probabilistic approach to related combinatorial challenges, such as the Maximum Happy Edges (MHE) problem or community detection in multi-layer networks, represents a promising direction for broader applicability in complex systems analysis.

\section*{Declarations of interest} 
None

\bibliographystyle{unsrt}
\bibliography{HC.bib}
\end{document}